\documentclass[a4paper,11pt]{article}
\usepackage{amssymb,amsmath,amsfonts,placeins,pbox,multirow,mathtools,bbold}
\usepackage[pdftex]{graphicx}
\usepackage{rotate,color,slashed,cite,epstopdf,verbatim,url,multirow,booktabs}
\usepackage{epstopdf}
\usepackage{subfig}
\usepackage{graphicx}
\usepackage{changepage}
\usepackage[normalem]{ulem}
\usepackage{minted}
\usepackage{cancel}
\graphicspath{{./Figures/}}
\usepackage{array}
 
\usepackage[a4paper, left=2cm, right=2cm, top=2cm, bottom=2cm]{geometry}
\usepackage{axodraw2}
\usepackage{epsfig}

\usepackage[utf8x]{inputenc}

\usepackage[dvipsnames]{xcolor}
\usepackage{axodraw2}
\usepackage[colorlinks=true,
            linkcolor=magenta,
            urlcolor=blue,
            citecolor=blue]{hyperref}
\usepackage[capitalise]{cleveref}

\crefname{section}{Sec.}{Secs.}
\crefname{table}{Tab.}{Tabs.}
\crefname{figure}{Fig.}{Figs.}
\crefname{equation}{Eq.}{Eqs.}
\crefname{appendix}{Appendix}{Appendix}
\makeatletter
\let\oldappendix\appendix
\renewcommand{\appendix}{%
  \oldappendix
  \crefalias{section}{appendix}%
}
\makeatother
\numberwithin{equation}{section}

\newcommand{\MET}{\ensuremath{E_T^\mathrm{miss}}}
\newcommand{\Lag}{\mathcal L}

\def\ssbh#1//#2//{\ensuremath{\xrightarrow [\substack{#2}]
    {\parbox{3cm}{\hfil $\scriptstyle \langle #1 \rangle$ \hfil}}}}

\let\CapTion=\caption
\def\caption#1{\CapTion{\em #1}}

\usepackage{titlesec}

\titleclass{\subsubsubsection}{straight}[\subsubsection]
\newcounter{subsubsubsection}[subsubsection]
\renewcommand\thesubsubsubsection{\thesubsubsection.\arabic{subsubsubsection}}
\titleformat{\subsubsubsection}
  {\normalfont\normalsize\bfseries}
  {\thesubsubsubsection}{1em}{}
\titlespacing*{\subsubsubsection}{0pt}{3.25ex plus 1ex minus .2ex}{1.5ex plus .2ex}

\makeatletter
\def\toclevel@subsubsubsection{4}
\def\l@subsubsubsection{\@dottedtocline{4}{7em}{4em}}
\makeatother

\title{\bf Dark matter in the x-channel}

\author{
Luca Panizzi\thanks{luca.panizzi@unical.it}
\\ \medskip 
\small\em Dipartimento di Fisica, Universit\`a della Calabria, I-87036 Arcavacata di Rende, Cosenza, Italy
}
\date{}

\begin{document}

\maketitle

% % % % % % % % % % % % % % % % Abstract % % % % % % % % % % % % % % % % % 

\begin{abstract}
In this paper I study a class of scenarios in which a scalar singlet dark matter candidate interacts with the Standard Model in such a way that the Higgs portal is strongly suppressed, even at loop level, through a specific mechanism involving scalar mediators. 
The mediator carries non-trivial gauge or internal quantum numbers that allow only a quartic coupling with the dark matter, and belongs to representations with no components that can mix with the Higgs boson, so that the dark matter interacts only with exotic states.
I refer to this construction as \emph{$x$-channel} because no single mediator is exchanged between the dark and visible sectors, to distinguish it from the standard $s$- and $t$-channel simplified scenarios. 
The suppression is due to a boundary condition on the running Higgs-portal coupling, which makes the tree-level and one-loop Higgs-mediated contributions vanish at zero momentum transfer, regardless of the internal quantum numbers of the mediator.
I consider here a minimal realisation, in which the mediator is a singly- or doubly-charged colourless $SU(2)_L$-singlet scalar decaying to leptons, and study the complementarity between collider, Higgs, flavour, and astrophysical/cosmological constraints. 
Within the scanned ranges and benchmark assumptions, the surviving points are predominantly in the region where the dark matter is heavier than the charged scalar mediator. LHC limits determine the lower edge of this region, part of which can be probed by future colliders through the production of the charged scalars. Other realisations of the $x$-channel can open different phenomenological directions to explore.
\end{abstract}

% % % % % % % % % % % % % % % Main Document% % % % % % % % % % % % % % % % 
\newpage
\tableofcontents
\bigskip
\hrule
\bigskip
% % % % % % % % % % % % % % % % % % % % % % % % % % % % % % % % % % % % % %

%%%%%%%%%%%%%%%%%%%%%%%%%%%%%%%%%%%%%%%%%%%%%%%%%%%%%%%%%%%%%%%%%%%%%%%%%%%%%%
\section{Introduction}
\label{sec:intro}

The nature of dark matter (DM) is one of the main open questions in particle physics and cosmology. Given the large parameter space of full theoretical models (see for example~\cite{Cirelli:2024ssz} for a review) and the need to find common features which can be tested in experiments, it has become standard to study simplified scenarios. The simplest of these is a scalar singlet $X_s$ interacting with the Standard Model (SM) through the Higgs portal~\cite{Silveira:1985ij,McDonald:1993ex,Burgess:2000yq,Cline:2013gha}, but its parameter space is heavily constrained~\cite{Goncalves:2025snm}. 
Introducing a mediator can help in opening up the available mass range at the price of introducing more free parameters. Non-minimal extensions involving a mediator are broadly categorised according to how such mediator transforms under the symmetry which stabilises the dark matter candidate, usually a $\mathbb{Z}_2$ parity under which the DM transforms oddly: in $s$-channel simplified scenarios  the mediator transforms evenly, so that it can couple directly to two DM states, while in $t$-channel scenarios it transforms oddly, so that any interaction must involve an even number of DM and mediator states overall. 

For the singlet dark matter case, a different mechanism can be obtained if the mediator is a scalar $\phi$ carrying a conserved internal quantum number, such as electric charge or colour, under which the DM is neutral. If the mediator is $\mathbb{Z}_2$-even, as in $s$-channel scenarios, a trilinear coupling with two DM states would still violate the conservation of the internal quantum number. The DM particle can however interact with $\phi$ through the renormalisable quartic coupling $X_s^2\phi^\dagger\phi$. If $\phi$ belongs to a representation with no components that can mix with the Higgs boson, the DM can only couple to exotic states with no other Higgs portal source. The mediator then communicates with the SM through its gauge interactions and, depending on the model, through couplings to the Higgs sector and to SM fermions. Unlike in the usual $s$- and $t$-channel simplified models, no single mediator is exchanged between the DM and visible sectors. 

Charged-scalar portals, in which the mediator is electrically charged, are a minimal realisation of such scenarios. They have been explored in connection with the tentative $130$~GeV gamma-ray excess in Fermi-LAT data~\cite{Bringmann:2012vr,Weniger:2012tx}, which later disappeared~\cite{Fermi-LAT:2013thd,Fermi-LAT:2015kyq}, where a real scalar DM particle couples through a quartic interaction to a charged scalar $S$ carrying QCD or hidden-$SU(N)$ charge~\cite{Cline:2012nw}. Subsequent studies considered charged scalar mediators within the Zee-Babu model~\cite{Baek:2012ub} and the Higgs Triplet Model~\cite{Wang:2012ts}. In~\cite{Hierro:2016nwm} a real singlet DM candidate couples to an $SU(2)_L$-singlet doubly-charged scalar, with the doubly-charged scalar also playing a role in neutrino masses and mixings. A very recent study also considers this setup~\cite{Borah:2026aqv} (see the note added after the Conclusions). All these examples still contain a direct Higgs-portal interaction of the scalar DM.

The quartic interactions of the mediator with the DM and with the Higgs boson also generate an effective coupling between the DM and the Higgs boson at loop level, which corrects the tree-level Higgs-portal coupling. 
It would be interesting to understand whether the Higgs portal can be suppressed almost completely, at least at the energy scales relevant for the observables, in the same fashion as~\cite{Belyaev:2022zjx,Belyaev:2022shr}. In this case the DM can talk to the visible sector only through the interactions of the mediator with the other SM fields, which then determine the observable consequences. I therefore explore a region in which both the tree-level and the loop-level Higgs portals are tiny or absent. 
I refer to the resulting mechanism as the {\it $x$-channel}, from the shape of the 4-leg topology which drives the phenomenology of this scenario.
Concretely, the $x$-channel regime can be defined by the boundary condition $\lambda_4(\mu_0)=0$ on the running coefficient of the Higgs-portal operator $\lambda_4 (H^\dagger H) X_s^2$, with $\mu_0=m_\phi$. The condition sets the Higgs-portal coefficient to zero at the mediator scale and, for this choice of $\mu_0$, the one-loop Higgs form factor vanishes at zero momentum transfer ($q^2=0$).

Since the $x$-channel condition involves only the mass of the mediator and its quartic couplings to the dark matter and to the Higgs boson, it does not depend on the internal quantum numbers of the mediator. Scalars with fractional electric charge, with colour charge, or in a non-trivial $SU(2)_L$ representation can therefore realise the $x$-channel. The mediator can also carry a charge under a new gauge, global or discrete symmetry. If this symmetry is exact and no operator allows $\phi$ to decay, $\phi$ would provide an additional stable relic. I do not consider such cases here and restrict the analysis to mediators that can decay to Standard Model states.

In this work I study a minimal realisation, in which the mediator is a colourless $SU(2)_L$ singlet. Requiring renormalisable interactions, it decays to leptons and can only be singly- or doubly-charged. Higher-charge singlets also realise the $x$-channel, but need higher-order operators or further mediators to decay into the SM, and are not treated here. I consider the combination of collider, Higgs, flavour, relic density, direct and indirect detection constraints for both singly- and doubly-charged mediators. For this realisation the LHC bounds on the pair production of the charged scalars dominate. The surviving points lie predominantly at $m_{X_s}\gtrsim m_\phi$ and extend up to $m_\phi\sim10$~TeV, and part of this region can be tested at future colliders.

The paper is organised as follows. \Cref{sec:xchannel} introduces the $x$-channel mechanism and the $x$-channel condition for a generic mediator. \Cref{sec:param} defines a minimal realisation with singly- and doubly-charged singlets, whose collider, Higgs, flavour, and cosmological and astrophysical constraints are discussed in \cref{sec:constraints}. In \cref{sec:nonminimal}, an overview of possible non-minimal scenarios which can accommodate the $x$-channel mechanism is given. The conclusions are given in \cref{sec:conclusion}. The renormalisation of the Higgs-portal operator and the loop amplitudes for the relic density are collected in \cref{app:renorm-derivation} and \cref{app:relicloops}, respectively.

%%%%%%%%%%%%%%%%%%%%%%%%%%%%%%%%%%%%%%%%%%%%%%%%%%%%%%%%%%%%%%%%%%%%%%%%%%%%%%
\section{The $x$-channel mechanism}
\label{sec:xchannel}

The mechanism involves two new fields: a scalar singlet $X_s$, which is the dark matter candidate, and a scalar $\phi$, which is the mediator. The dark matter scalar is assumed to be real in this study; extending to a complex scalar is straightforward, would modify the observable predictions, but not the main mechanism. The Lagrangian of $X_s$ is given by:
\begin{equation}
\label{eq:LagX}
\Lag_{X_s} = {1\over2} (\partial_\mu X_s)^2 - {1\over2}m_{X_s}^2 X_s^2 - {\lambda_X\over4} X_s^4 - {\lambda_4\over2} H^2 X_s^2\;,
\end{equation}
where $H$ is the SM Higgs doublet. The renormalisable operator $\mu_3 H^2 X_s$ would make $X_s$ unstable, and therefore not a possible DM candidate. It is forbidden by imposing a $\mathbb Z_2$ parity under which $X_s$ is the only particle transforming oddly.\footnote{In case of a complex scalar $X_c$, this condition would be equivalent to the conservation of the internal quantum number carried by the complex scalar.}
Parameters are chosen such that the electroweak vacuum preserves this $\mathbb Z_2$ parity, so $X_s$ has vanishing VEV and cannot mix with the Higgs boson.

The mediator $\phi$ is $\mathbb{Z}_2$-even and carries a conserved internal quantum number, such as electric charge or colour, under which $X_s$ is neutral. Its representation is required to contain no components that can mix with the Higgs boson, to avoid introducing a Higgs portal through other particles. This limits the couplings of the dark matter to exotic states. The mediator $\phi$ can couple to $X_s$ only through the combination $X_s^2\phi^\dagger\phi$ ($X_s^2\phi^a\phi^a$ for a real representation). The couplings of $\phi$ to the dark matter and to the Higgs doublet relevant for this mechanism are therefore quartic,
\begin{equation}
\label{eq:Lagphigen}
\Lag_\phi \supset {\kappa_4 \over 2} (\phi^\dagger \phi) X_s^2 + \kappa_H (H^\dagger H) (\phi^\dagger \phi)\;.
\end{equation}

\subsection{The Higgs portal and the $x$-channel condition}
\label{app:renorm}

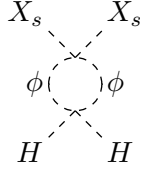
\begin{figure}[h]
\centering
\begin{picture}(70,56)(0,3)
\Line[dash](30,50)(40,40)
\Text(28,52)[br]{$X_s$}
\Line[dash](50,50)(40,40)
\Text(52,52)[bl]{$X_s$}
\Arc[dash,clock](40,30)(10,180,0)
\Text(52,30)[lc]{$\phi$}
\Arc[dash,clock](40,30)(10,0,180)
\Text(28,30)[rc]{$\phi$}
\Line[dash](30,10)(40,20)
\Text(28,8)[tr]{$H$}
\Line[dash](50,10)(40,20)
\Text(52,8)[tl]{$H$}
\end{picture}
\caption{\label{fig:H2X2operator}
The radiatively-generated Higgs portal.
}
\end{figure}

The initial working assumption is that the tree-level Higgs-portal coupling $\lambda_4$ is small enough to give unobservable contributions.
The operator $(H^\dagger H) X_s^2$, however, is allowed by all symmetries of \cref{eq:LagX}. Therefore, even in the absence of the tree-level $\lambda_4$ term, an effective quartic coupling is generated radiatively by the diagram in \cref{fig:H2X2operator}. 

A stronger condition to limit the effect of the Higgs portal is therefore given by imposing that the running quartic coupling vanishes at a scale which is relevant for the loop contribution
\begin{equation}
\label{eq:xchannelcondition}
\lambda_4(\mu_0)=0\,,
\end{equation}
where since the only particle circulating in the one-loop diagrams is $\phi$, it is natural to associate the reference scale to its mass:
\begin{equation}
\label{eq:mu0definition}
\mu_0=c\,m_\phi\,,
\end{equation}
with $c$ a dimensionless parameter of $\mathcal{O}(1)$. \Cref{eq:xchannelcondition,eq:mu0definition} define the $x$-channel condition.

Renormalising in the $\overline{\rm MS}$ scheme and imposing \cref{eq:xchannelcondition}, the $\mu$-independent bubble contribution to the $(H^\dagger H) X_s^2$ vertex is (see \cref{app:renorm-derivation})
\begin{equation}
\label{eq:lam4eff}
\lambda_{4,\rm eff}(q^2;\mu_0)
=
-\frac{\kappa_4\kappa_H}{16\pi^2}
\left[
f\!\left(\frac{q^2}{m_\phi^2}\right)
+
2\ln c
\right]\,,
\end{equation}
where $f$ is the $\mu$-independent part of the equal-mass $B_0$ function~\cite{Denner:2019vbn},
\begin{equation}
\label{eq:fdef}
f(x) \equiv 2 - \frac{1}{x}\left(\frac{1}{r}-r\right)\ln r\,,\qquad r+\frac{1}{r}=2-x\,.
\end{equation}
For $x\ll1$, $f$ expands as
\begin{equation}
\label{eq:fsmallx}
f(x) = \frac{x}{6} + \frac{x^2}{60} + \mathcal{O}(x^3)\,,
\end{equation}
so that in the $q^2\to0$ or $m_\phi\to\infty$ limits $f(q^2/m_\phi^2)$ vanishes and
\begin{equation}
\label{eq:lam4effzero}
\lambda_{4,\rm eff}(0;c)
=
-\frac{\kappa_4\kappa_H}{8\pi^2}
\ln c\,.
\end{equation}
The choice $c=1$ makes the Higgs portal vanish at one-loop in this limit and is retained throughout the paper, as it is the unique choice for which the renormalised zero-momentum form factor vanishes for $\kappa_4\kappa_H\neq0$.

In the rest of the paper, $\kappa_H$ and $\kappa_4$ are treated as scale-independent inputs.
A fully self-consistent one-loop treatment would give $\kappa_H$ and $\kappa_4$ their own RGEs. Dropping them is consistent provided $\kappa_H,\kappa_4$ vary little between $\mu_0=c\,m_\phi$ and the scales entering freeze-out, (in)direct detection and collider observables.

%%%%%%%%%%%%%%%%%%%%%%%%%%%%%%%%%%%%%%%%%%%%%%%%%%%%%%%%%%%%%%%%%%%%%%%%%%%%%%
\section{A minimal realisation with electrically charged singlets}
\label{sec:param}

A minimal realisation of the $x$-channel is obtained when $\phi$ is a colourless, $U(1)_Y$-charged $SU(2)_L$ singlet. Requiring renormalisable interactions, it decays to leptons and can only be singly- or doubly-charged electrically, $\phi^{+(+)}$. For simplicity, the singly- and doubly-charged scenarios are treated separately. Extending the treatment to a combined analysis of the charged scalars or considering different mediators is left for a future analysis. The Lagrangian of the singly-charged scalar is:
\begin{eqnarray}
\Lag_{\phi^+} &=& D_\mu \phi^+ D^\mu \phi^- - \mu_\phi^2 \phi^+\phi^- \nonumber\\
&+& {\kappa_4 \over 2} (\phi^+ \phi^-) X_s^2 + \kappa_H (H^\dagger H) (\phi^+ \phi^-) - \left( {\lambda_{ij}\over 2} \varepsilon^{ab} \bar{L}_{ia}^c L_{jb} \phi^+ + {\rm h.c.} \right)\;,
\label{eq:Lagphip}
\end{eqnarray}
where $L$ are SM lepton doublets, the $c$ superscript represents the charge conjugation operator, $i,j$ are indices in flavour space, $a,b$ are indices of $SU(2)$ and $\varepsilon^{ab}$ is the $2\times2$ antisymmetric tensor. The matrix $\lambda_{ij}$ is thus antisymmetric and the singly-charged scalar can only couple to charged leptons and neutrinos of different flavours ({\it e.g.} $\lambda_{e\mu}~(\bar\nu_e^c \mu_L - \bar \nu_\mu^c e_L)\phi^+$). The Lagrangian of the doubly-charged scalar is:
\begin{eqnarray}
\Lag_{\phi^{++}} &=& D_\mu \phi^{++} D^\mu \phi^{--} - \mu_\phi^2 \phi^{++}\phi^{--} \nonumber\\
&+& {\kappa_4 \over 2} (\phi^{++} \phi^{--}) X_s^2 + \kappa_H (H^\dagger H) (\phi^{++} \phi^{--}) - \left( f_{ij} \bar \ell_{iR}^c \ell_{jR} \phi^{++} + {\rm h.c.}\right)\;,
\label{eq:Lagphipp}
\end{eqnarray}
where $\ell_R$ are SM right-handed leptons and, differently from the singly-charged case, $f_{ij}$ is symmetric and $\phi^{++}$ can couple to charged leptons of same or different flavour.

After the Higgs acquires its VEV, the $\phi^+$ and $\phi^{++}$ masses are given by the same relation
\begin{equation}
m_\phi^2 \equiv \mu_\phi^2 - \kappa_H v^2/2 \quad\text{with}\quad \phi=\phi^+,\phi^{++}\;,
\end{equation}
and $m_\phi$ will be treated as an independent input parameter in the following. From now on, when there is no ambiguity, $\phi^+$ and $\phi^{++}$ will be collectively labelled as $\phi$. 

Notice also that, limiting the analysis to renormalisable operators, at least one leptonic decay must be allowed for $\phi$, otherwise it would be stable too and inconsistent with the observation of a neutral Universe. The displayed Lagrangians contain only the operators relevant to the observables studied here; in particular, no self-quartic $\lambda_\phi(\phi^\dagger\phi)^2$ is included.

%%%%%%%%%%%%%%%%%%%%%%%%%%%%%%%%%%%%%%%%%%%%%%%%%%%%%%%%%%%%%%%%%%%%%%%%%%%%%%
\section{Constraints on the minimal scenario}
\label{sec:constraints}

The simultaneous presence of a dark matter candidate and of an electrically charged mediator interacting with SM leptons allows testing the minimal $x$-channel scenarios from multiple directions, involving cosmological and astrophysical probes, flavour physics observables and collider processes. Other realisations of the $x$-channel have production and decay patterns set by the quantum numbers of the mediator, leading to completely different phenomenological consequences.

The free parameters of the minimal model are the following:
\begin{equation}
m_{X_s},~m_\phi,~\kappa_4,~\kappa_H, \text{~and~}\lambda_{ij} \text{~or~} f_{ij} \text{~depending on the charge of~}\phi\;.
\end{equation}
The allowed parameter space of the model has been determined by performing a scan over the free parameters in logarithmic scale. Masses of DM and charged scalars have been varied in the interval $\{0.1,10^5\}$ GeV, the quartic coupling $\kappa_4$ in the interval $\{10^{-5},4\pi\}$, the portal coupling between the charged scalars and the Higgs boson $\kappa_H$ and all leptonic couplings in the interval $\{10^{-9},4\pi\}$. The higher minimum value of $\kappa_4$ avoids points that always predict an overabundant relic density, while the lower minimum allowed for $\kappa_H$ and the leptonic couplings makes it possible to probe regimes in which all Higgs-mediated processes are negligible. Due to the many possible forms of the lepton coupling matrices, the numerical scan assumes universal leptonic couplings for the charged and doubly-charged scalars, including lepton-flavour-violating terms for $\phi^{++}$. 

\subsection{Collider and Higgs bounds} 
\label{sec:LHC}
The strongest constraint on the parameter space of the model comes from Drell-Yan (DY) pair production of charged scalars at colliders, through $s$-channel photon/$Z$ exchange.

The only dedicated searches for singly-charged scalars date back to LEP, under the hypothesis that $\phi^+$ belongs to a doublet in a 2HDM extension of the SM, and therefore targeted predominantly the $\tau\nu$ decay~\cite{ALEPH:2013htx}. Bounds were reinterpreted for different isospin values in~\cite{Banerjee:2025qkb}, setting an upper limit on the $\tau\nu$ branching ratio as function of $m_\phi$ in the range 45 GeV $\lesssim m_\phi \lesssim$ 95 GeV. However, $\phi^+$ shares the gauge quantum numbers of a right-handed slepton. At the LHC, because the antisymmetry of $\lambda_{ij}$ forces lepton-flavour-violating decays $\phi^+\to\ell_i^+\nu_j$ with $i\neq j$, the signal topology differs from the same-flavour assumption underlying standard slepton searches; a recast of an ATLAS slepton analysis~\cite{ATLAS:2019lff} yields a coupling-independent lower bound of approximately $200$~GeV~\cite{Crivellin:2020klg}, which is the conservative value used in the following. A dedicated recent study of $SU(2)_L$- and $SU(3)_c$-singlet singly-charged scalars decaying promptly to a lepton plus an invisible state~\cite{Das:2025qkb} finds surviving mass windows below and above the region excluded by existing slepton searches, depending on the assumed branching ratios.

Doubly-charged scalars have been explored more consistently in experimental searches. The ATLAS search~\cite{ATLAS:2022pbd} looks for final states with two, three or four light leptons and, under the assumption of equal branching ratios into any lepton combination, including flavour-violating ones, sets $m_{\phi^{++}}>900$ GeV for a singlet doubly-charged scalar in the Zee-Babu benchmark. I adopt this value as the benchmark collider limit. The numerical bound is model-dependent and would change for different flavour structures. 

\begin{figure}[h!]
\centering
\begin{picture}(150,50)(0,0)
\Line[](10,40)(40,20)
\Text(8,39)[rc]{$q$}
\Line[](10,00)(40,20)
\Text(8,-1)[rc]{$\bar q$}
\Photon(40,20)(70,20){2}{3}
\Text(55,24)[cb]{$\gamma,Z$}
\Line[dash](70,20)(90,35)
\Text(79,33)[cb]{$\phi^+$}
\Line[dash](70,20)(90,5)
\Text(79,8)[ct]{$\phi^+$}
\Line[dash](90,35)(90,5)
\Text(92,20)[cl]{$\phi^+$}
\Line[dash](90,35)(110,40)
\Text(112,40)[lc]{$X_s$}
\Line[dash](90,35)(110,30)
\Text(112,30)[lc]{$X_s$}
\Line[dash](90,5)(110,0)
\Photon(90,5)(110,0){2}{3}
\Text(112,0)[cl]{$\gamma,Z,h$}
\end{picture}
\begin{picture}(150,50)(0,0)
\Line[](10,40)(40,20)
\Text(8,39)[rc]{$q$}
\Line[](10,00)(40,20)
\Text(8,-1)[rc]{$\bar q$}
\Photon(40,20)(70,20){2}{3}
\Text(55,25)[cb]{$Z$}
\Photon(70,20)(90,0){2}{3}
\Text(90,-2)[cl]{$Z$}
\Line[dash](70,20)(80,25)
\Text(72,24)[cb]{\rotatebox{35}{$h$}}
\Arc[dash,clock](90,29)(10,0,360)
\Text(89,38)[cb]{\rotatebox{35}{$\phi^{+(+)}$}}
\Text(103,30)[ct]{\rotatebox{35}{$\phi^{+(+)}$}}
\Line[dash](100,35)(110,45)
\Text(111,48)[lc]{$X_s$}
\Line[dash](100,35)(112,35)
\Text(114,35)[lc]{$X_s$}
\end{picture}\\
\begin{picture}(140,100)(0,0)
\Line[](10,70)(40,70)
\Text(8,69)[rc]{$q$}
\Line[](10,10)(40,10)
\Text(8,9)[rc]{$q$}
\Line[](40,70)(100,80)
\Text(102,79)[lc]{$q$}
\Line[](40,10)(100,0)
\Text(102,-1)[lc]{$q$}
\Photon(40,70)(50,50){2}{3}
\Text(40,61)[rt]{$\gamma,Z$}
\Photon(40,10)(50,30){2}{3}
\Text(40,17)[rb]{$\gamma,Z$}
\Line[dash](50,30)(50,50)
\Text(48,40)[rc]{$\phi^+$}
\Line[dash](50,50)(70,40)
\Text(60,49)[lb]{$\phi^+$}
\Line[dash](70,40)(50,30)
\Text(60,33)[lt]{$\phi^+$}
\Line[dash](70,40)(100,55)
\Text(102,55)[lc]{$X_s$}
\Line[dash](100,25)(70,40)
\Text(102,25)[lc]{$X_s$}
\end{picture}
\begin{picture}(150,80)(0,-10)
\Line[](10,50)(40,30)
\Text(8,49)[rc]{$q$}
\Line[](10,10)(40,30)
\Text(8,9)[rc]{$\bar q$}
\Photon(40,30)(60,30){2}{3}
\Text(50,35)[cb]{$Z$}
\Line[](60,30)(80,45)
\Text(67,40)[cb]{$\nu_i$}
\Line[](60,30)(80,15)
\Text(67,19)[ct]{$\bar \nu_i$}
\Line[](80,45)(120,60)
\Text(122,60)[lc]{$\ell_j^-$}
\Line[](80,15)(120,0)
\Text(122,0)[lc]{$\ell_k^+$}
\Line[dash](80,45)(100,30)
\Text(96,40)[cb]{$\phi^+$}
\Line[dash](100,30)(80,15)
\Text(96,23)[ct]{$\phi^+$}
\Line[dash](100,30)(120,40)
\Text(122,40)[lc]{$X_s$}
\Line[dash](100,30)(120,20)
\Text(122,20)[lc]{$X_s$}
\end{picture}
\begin{picture}(150,80)(0,-10)
\Line[](10,50)(40,30)
\Text(8,49)[rc]{$q$}
\Line[](10,10)(40,30)
\Text(8,9)[rc]{$\bar q$}
\Photon(40,30)(60,30){2}{3}
\Text(49,34)[cb]{$\gamma,Z$}
\Line[](60,30)(80,45)
\Text(67,40)[cb]{$\ell^-_i$}
\Line[](60,30)(80,15)
\Text(67,19)[ct]{$\ell^+_i$}
\Line[](80,45)(120,60)
\Text(122,60)[lc]{$\ell^+_j$}
\Line[](80,15)(120,0)
\Text(122,0)[lc]{$\ell^-_k$}
\Line[dash](100,30)(80,45)
\Text(98,40)[cb]{$\phi^{++}$}
\Line[dash](80,15)(100,30)
\Text(98,23)[ct]{$\phi^{++}$}
\Line[dash](100,30)(120,40)
\Text(122,40)[lc]{$X_s$}
\Line[dash](100,30)(120,20)
\Text(122,20)[lc]{$X_s$}
\end{picture}

\caption{\label{fig:LHCtopologies} Representative topologies for $x$-channel DM production at the LHC. {\bf Top row:} mono-$\gamma,Z,h$. {\bf Bottom row:} di-jet+$\MET$ on the left, and 2$\ell$+$\MET$ mediated by $\phi^+$ and $\phi^{++}$ on the centre and right panels, respectively.}
\end{figure}
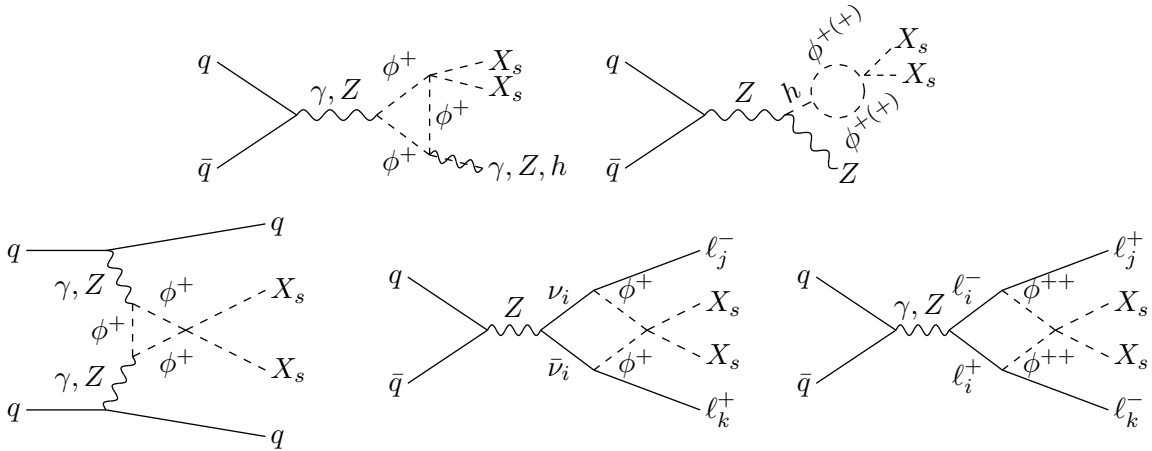

Direct production of the dark matter candidate at the LHC is necessarily loop-induced, since $X_s$ couples only to the charged scalar $\phi$, which carries no colour charge and has no direct coupling to initial-state partons. The processes through which $X_s$ can be produced at the LHC are shown in \Cref{fig:LHCtopologies}. 

Mono-jet topologies can arise only through Higgs-mediated diagrams, in which the Higgs couples to the initial-state quarks through their Yukawa interactions and to the dark sector through a $\phi$ loop, with the jet radiated from an initial-state parton, and this process is therefore negligible. Exchange of a single electroweak gauge boson does not contribute: the one-loop $X_sX_s\gamma$ and $X_sX_sZ$ vertices both vanish by charge-conjugation invariance. The mono-$\gamma$, mono-$Z$ and mono-$h$ final states remain open, with the SM boson emitted either directly from the $\phi$ loop or, for mono-$Z$, from a diagram in which an off-shell Higgs radiated by the $\phi$ loop splits into $Zh$ before the Higgs converts to the $X_sX_s$ pair.

The model is implemented in UFO format~\cite{Degrande:2011ua,Darme:2023jdn} via {\sc FeynRules}~\cite{Alloul:2013bka}, with one-loop counterterms from {\sc NLOCT}~\cite{Degrande:2014vpa}, manually edited to include finite terms for the effective coupling between Higgs and dark matter. Mono-$X$ events are generated with {\sc MG5\_aMC}~\cite{Alwall:2014hca} using the {\sc NNPDF40\_lo\_as\_01180} parton densities~\cite{NNPDF:2021njg} through {\sc LHAPDF}~\cite{Buckley:2014ana}, showered with {\sc Pythia8}~\cite{Sjostrand:2014zea}, and recast with {\sc MadAnalysis5}~\cite{Conte:2018vmg} against ATLAS and CMS 13~TeV searches for missing transverse energy (\MET) plus one or more SM objects.

The exclusion bounds are dominated, for both $\phi^+$ and $\phi^{++}$, by the mono-$Z$ process, in particular through the topology with Higgs exchange (top-right in \cref{fig:LHCtopologies}). Mono-$\gamma$ and mono-$h$ proceed only through the triangle topology, with cross sections below 1~ab, far from the reach of the LHC and HL-LHC after accounting for experimental efficiencies.

\begin{figure}[h]
\centering
\epsfig{file=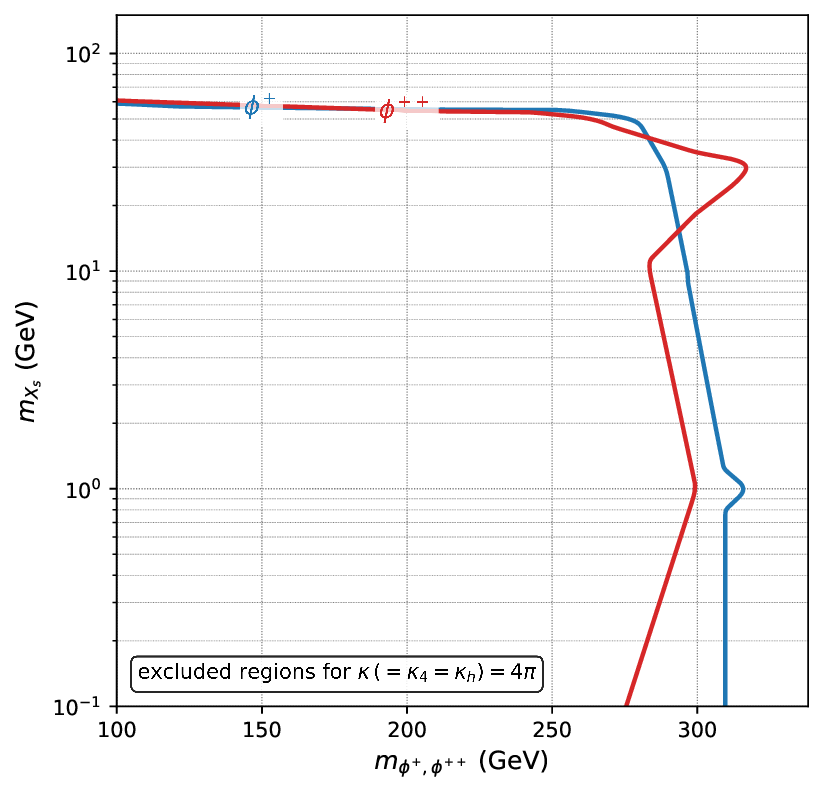,width=.49\textwidth}\hfill
\epsfig{file=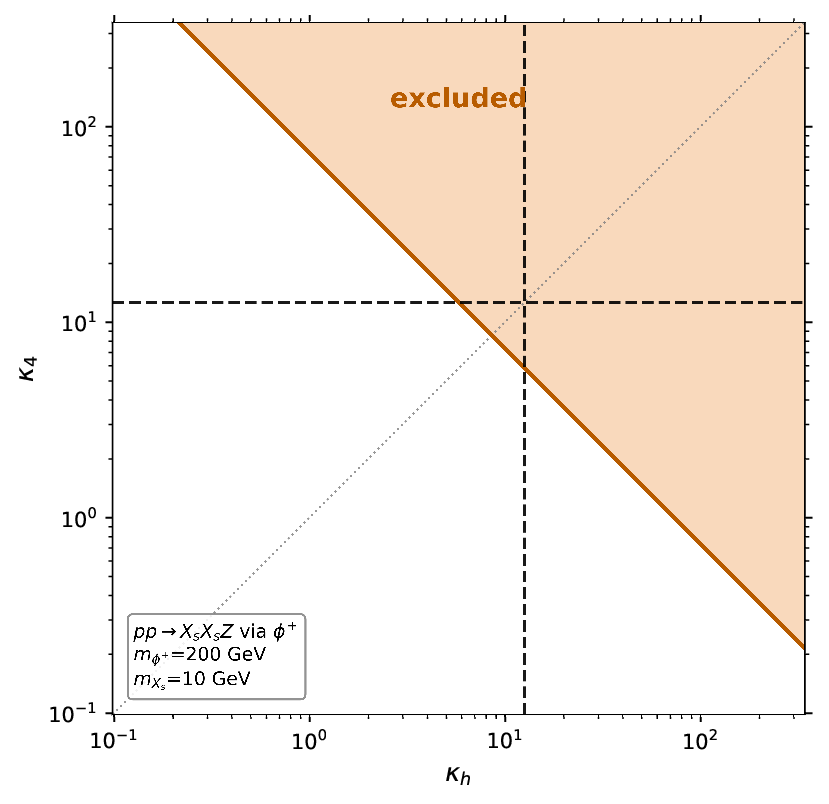,width=.49\textwidth}
\caption{\label{fig:LHCrecast} Recast bound from mono-$Z$ for a scenario in which $\kappa_4=\kappa_H=4\pi$ (left) and as function of the two separate couplings (right) for the $\phi^+$ case, with $m_{\phi^+}=200$ GeV and $m_{X_s}=10$ GeV.}
\end{figure}

The mono-$Z$ bound depends on both $\kappa_4$ and $\kappa_H$. 
The left panel of \cref{fig:LHCrecast} shows the bound for $\kappa_4=\kappa_H=4\pi$, the upper edge of the scan.
Since the interactions involved do not depend on the $\phi$ charge, the bounds are the same for $\phi^+$ and $\phi^{++}$, barring fluctuations. The excluded region reaches $m_{X_s}\simeq m_h/2$, where the Higgs boson decays into the DM pair through the $\phi$ loop, for $m_\phi\lesssim300$~GeV, and disappears rapidly above this mass because of the loop suppression. The right panel shows the dependence on the two couplings for representative masses: the amplitude squared is quadratic in both couplings in the dominant topology, so the excluded region is symmetric under $\kappa_4\leftrightarrow\kappa_H$. Other mass configurations give analogous results.

The di-jet + $\MET$ final state can be obtained by multiple topologies including the vector-boson fusion one shown in the bottom-left of \cref{fig:LHCtopologies}; these proceed through the same loop-suppressed $\phi$-loop vertices as the mono-boson processes above, with an additional electroweak vertex and phase-space suppression from the two associated jets, and are not expected to compete with the mono-$Z$ bound. The 2$\ell$+$\MET$ final state has the same signature of slepton pair production, and therefore it is subdominant with respect to the $\phi$ pair production bounds obtained through tree level processes.

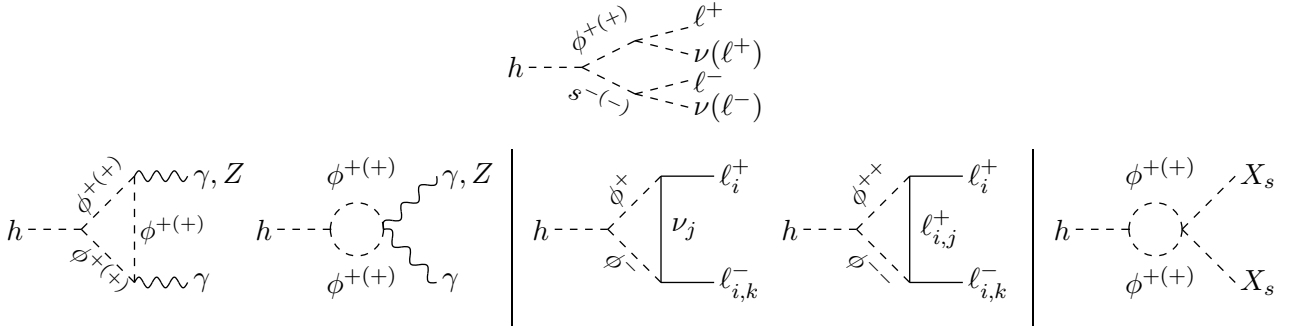
\begin{figure}[h!]
\centering
\begin{picture}(105,60)(0,0)
\Line[dash](10,30)(30,30)
\Text(8,30)[rc]{$h$}
\Line[dash](30,30)(50,40)
\Text(37,34)[cb]{\small\rotatebox{25}{$\phi^{+(+)}$}}
\Line[dash](30,30)(50,20)
\Text(37,27)[ct]{\small\rotatebox{-25}{$s^{-(-)}$}}
\Line[dash](50,40)(70,45)
\Text(72,50)[lc]{$\ell^+$}
\Line[dash](50,40)(70,35)
\Text(72,35)[lc]{$\nu (\ell^+)$}
\Line[dash](50,20)(70,25)
\Text(72,25)[lc]{$\ell^-$}
\Line[dash](50,20)(70,15)
\Text(72,15)[lc]{$\nu (\ell^-)$}
\end{picture}
\begin{tabular}{c|c|c}
\begin{picture}(90,50)(5,0)
\Line[dash](10,30)(30,30)
\Text(8,30)[rc]{$h$}
\Line[dash](30,30)(50,50)
\Text(37,34)[cb]{\small\rotatebox{45}{$\phi^{+(+)}$}}
\Line[dash](30,30)(50,10)
\Text(37,27)[ct]{\small\rotatebox{-45}{$\phi^{+(+)}$}}
\Line[dash](50,10)(50,50)
\Text(52,30)[cl]{\small$\phi^{+(+)}$}
\Photon(50,50)(70,50){2}{3}
\Text(72,50)[lc]{$\gamma,Z$}
\Photon(50,10)(70,10){2}{3}
\Text(72,10)[lc]{$\gamma$}
\end{picture}
\begin{picture}(87,50)(5,0)
\Line[dash](10,30)(30,30)
\Text(8,30)[rc]{$h$}
\Arc[dash,clock](40,30)(10,180,0)
\Text(42,45)[cb]{$\phi^{+(+)}$}
\Arc[dash,clock](40,30)(10,0,180)
\Text(42,16)[ct]{$\phi^{+(+)}$}
\Photon(50,30)(70,50){2}{3}
\Text(72,50)[lc]{$\gamma,Z$}
\Photon(50,30)(70,10){2}{3}
\Text(72,10)[lc]{$\gamma$}
\end{picture}
&
\begin{picture}(90,60)(0,0)
\Line[dash](10,30)(30,30)
\Text(8,30)[rc]{$h$}
\Line[dash](30,30)(50,50)
\Text(35,37)[cb]{\small\rotatebox{45}{$\phi^+$}}
\Line[dash](30,30)(50,10)
\Text(35,24)[ct]{\small\rotatebox{-45}{$\phi^-$}}
\Line(50,50)(50,10)
\Text(54,30)[cl]{$\nu_j$}
\Line(70,50)(50,50)
\Text(72,50)[lc]{$\ell^+_i$}
\Line(50,10)(70,10)
\Text(72,10)[lc]{$\ell^-_{i,k}$}
\end{picture} 
\begin{picture}(90,60)(0,0)
\Line[dash](10,30)(30,30)
\Text(8,30)[rc]{$h$}
\Line[dash](30,30)(50,50)
\Text(35,37)[cb]{\small\rotatebox{45}{$\phi^{++}$}}
\Line[dash](30,30)(50,10)
\Text(35,24)[ct]{\small\rotatebox{-45}{$\phi^{--}$}}
\Line(50,50)(50,10)
\Text(54,30)[cl]{$\ell^+_{i,j}$}
\Line(70,50)(50,50)
\Text(72,50)[lc]{$\ell^+_i$}
\Line(50,10)(70,10)
\Text(72,10)[lc]{$\ell^-_{i,k}$}
\end{picture}
&
\begin{picture}(95,50)(0,0)
\Line[dash](10,30)(30,30)
\Text(8,30)[rc]{$h$}
\Arc[dash,clock](40,30)(10,180,0)
\Text(42,45)[cb]{$\phi^{+(+)}$}
\Arc[dash,clock](40,30)(10,0,180)
\Text(42,16)[ct]{$\phi^{+(+)}$}
\Line[dash](50,30)(70,50)
\Text(72,50)[lc]{$X_s$}
\Line[dash](50,30)(70,10)
\Text(72,10)[lc]{$X_s$}
\end{picture}
\end{tabular}
\caption{\label{fig:hdecayloop} Contributions to Higgs decays in the $x$-channel. {\bf Top row:}
four-body decay $h\to\phi^{+(+)}\phi^{-(-)} \to 2\ell\,2\nu\;(4\ell)$.
{\bf Bottom row:} loop-induced $h\to\gamma\gamma,\,Z\gamma$ (left), loop-induced decays to leptons of same or different flavours (centre) and loop-induced invisible decay $h\to X_s X_s$ (right).}
\end{figure}

Higgs boson observables are affected by the charged scalar and by dark matter through the mechanisms shown in~\cref{fig:hdecayloop}. The resulting signal strengths and exclusion limits are evaluated using {\sc HiggsBounds} and {\sc HiggsSignals} via the {\sc HiggsTools} interface~\cite{Bahl:2022igd}. 

The tree-level decays $h\to\phi^+\phi^-\to2\ell\,2\nu$ and $h\to\phi^{++}\phi^{--}\to4\ell$ proceed through on-shell $\phi$ for $m_\phi<m_h/2$ and through off-shell propagators otherwise; given the collider bounds above, only the off-shell regime is relevant in practice. For $\phi^{++}$ the same-sign dilepton pairs are not accepted by the $ZZ^*\to4\ell$ selection, which requires opposite-sign same-flavour pairs compatible with the $Z$ and $Z^*$ masses, and the decay is treated as an undetected exotic width. For $\phi^+$ the $2\ell\,2\nu$ final state coincides with that of $WW^*$, but the lepton-neutrino invariant mass peaks near $m_\phi$: the partial width is attributed to the $WW$ channel of {\sc HiggsSignals} only for $m_\phi\simeq m_W$, and is otherwise treated as an undetected exotic width.

The loop-induced decays $h\to\gamma\gamma$ and $h\to Z\gamma$ receive contributions from two distinct topologies, and the coupling modifiers read:
\begin{eqnarray}
\label{eq:Hkmodifiers}
\kappa_{\gamma\gamma} \equiv
\frac{\bigl|A^{\gamma\gamma}_{\rm SM} - Q_\phi^2\,\dfrac{\kappa_H v^2}{2m_\phi^2}\,A_0(\tau_\phi)\bigr|}
     {\bigl|A^{\gamma\gamma}_{\rm SM}\bigr|}
\quad\text{and}\quad
\kappa_{Z\gamma} \equiv
\frac{\bigl|A^{Z\gamma}_{\rm SM} - 2Q_\phi^2\,\dfrac{s_w}{c_w}\,\dfrac{\kappa_H v}{m_\phi^2}\,A_0^{Z\gamma}(\tau_\phi,\lambda_\phi)\bigr|}
     {\bigl|A^{Z\gamma}_{\rm SM}\bigr|}\,,
\end{eqnarray}
where $A^{\gamma\gamma}_{\rm SM}=A_1(\tau_W)+N_cQ_t^2A_{1/2}(\tau_t)$, $\tau_i=4m_i^2/m_h^2$, $\lambda_i=4m_i^2/m_Z^2$, and $A_0$, $A_1$, $A_{1/2}$ are standard loop functions~\cite{Djouadi:2005gi,Chen:2013vi}. 

The loop-induced decay to leptons for the singly-charged scalar has been treated from a model-independent point of view in Ref.~\cite{Zeleny-Mora:2021tym}. Since $\phi^+$ interacts only with left-handed doublets and the internal neutrinos are massless, all these topologies give an identically zero contribution at leading order. Non-zero contributions arise only from the charged-lepton mass on the external legs, which makes the amplitude proportional to $m_\ell$, and this decay is neglected in the following.

For the doubly-charged scalar the amplitude is proportional to the mass of the external leptons,
\begin{equation}
A_L = X_{ij}\,m_{\ell_i}\,,\qquad A_R = X_{ij}\,m_{\ell_j}\,,\qquad
X_{ij}=\frac{4\kappa_H v}{16\pi^2}\sum_k f_{ik}f_{jk}\,C_1(m_{\ell_k}^2;m_\phi^2,m_\phi^2)\,,
\end{equation}
where $C_1$ is the Passarino-Veltman coefficient of the external momentum. 
For external masses much smaller than $m_h$ the flavour-violating width is
\begin{equation}
\Gamma(h\to\ell_i^\pm\ell_j^\mp)\simeq\frac{m_h}{8\pi}\,|X_{ij}|^2\left(m_{\ell_i}^2+m_{\ell_j}^2\right),\qquad i\neq j\,.
\end{equation}
For $i=j$ the amplitude is proportional to $m_\ell$ and interferes with the SM Yukawa coupling, $\Gamma(h\to\ell^+\ell^-)=\Gamma_{\rm SM}\,|1+\delta_\ell|^2$ with $\delta_\ell=vX_{\ell\ell}$.

The loop-induced invisible decay $h\to X_s X_s$ is open when $m_{X_s}<m_h/2$ and is mediated by a bubble of charged scalars. The amplitude follows directly from \cref{eq:lam4eff},
\begin{equation}
\label{eq:hxsxsamp}
\mathcal{M}(h\to X_s X_s) \;=\; -v\,\lambda_{4,\rm eff}(m_h^2)\;,
\end{equation}
leading to the partial width
\begin{equation}
\Gamma(h\to X_s X_s) \;=\; \frac{|\mathcal{M}|^2}{2}\cdot\frac{\beta_{X_s}}{16\pi\,m_h}\;,
\quad
\beta_{X_s} = \sqrt{1 - 4m_{X_s}^2/m_h^2}\;.
\end{equation}
This width is implemented in {\sc HiggsTools} as an undetected exotic decay. Since the dataset does not include a dedicated invisible Higgs measurement, a bound $\mathrm{BR}(h\to X_s X_s)<0.107$ is separately imposed, corresponding to the $95\%$~CL upper limit from the ATLAS combination~\cite{ATLAS:2023tkt}.

A point is excluded at $95\%$ CL when $\Delta\chi^2\equiv\chi^2_{x\text{-ch.}}-\chi^2_{\rm SM}$ exceeds $\chi^2_{0.95}(N_{\rm DoF})$, where $N_{\rm DoF}$ counts the parameters on which the Higgs observables depend: $m_\phi$ and $\kappa_H$ always enter through the loop-modified $h\to\gamma\gamma,Z\gamma$ rates (\cref{eq:Hkmodifiers}) and the $h\to\phi\phi^*$ width, $m_{X_s}$ and $\kappa_4$ are added when $h\to X_sX_s$ is open, and the lepton couplings when leptonic Higgs decays are sizeable.

\begin{figure}[h]
\centering
\epsfig{file=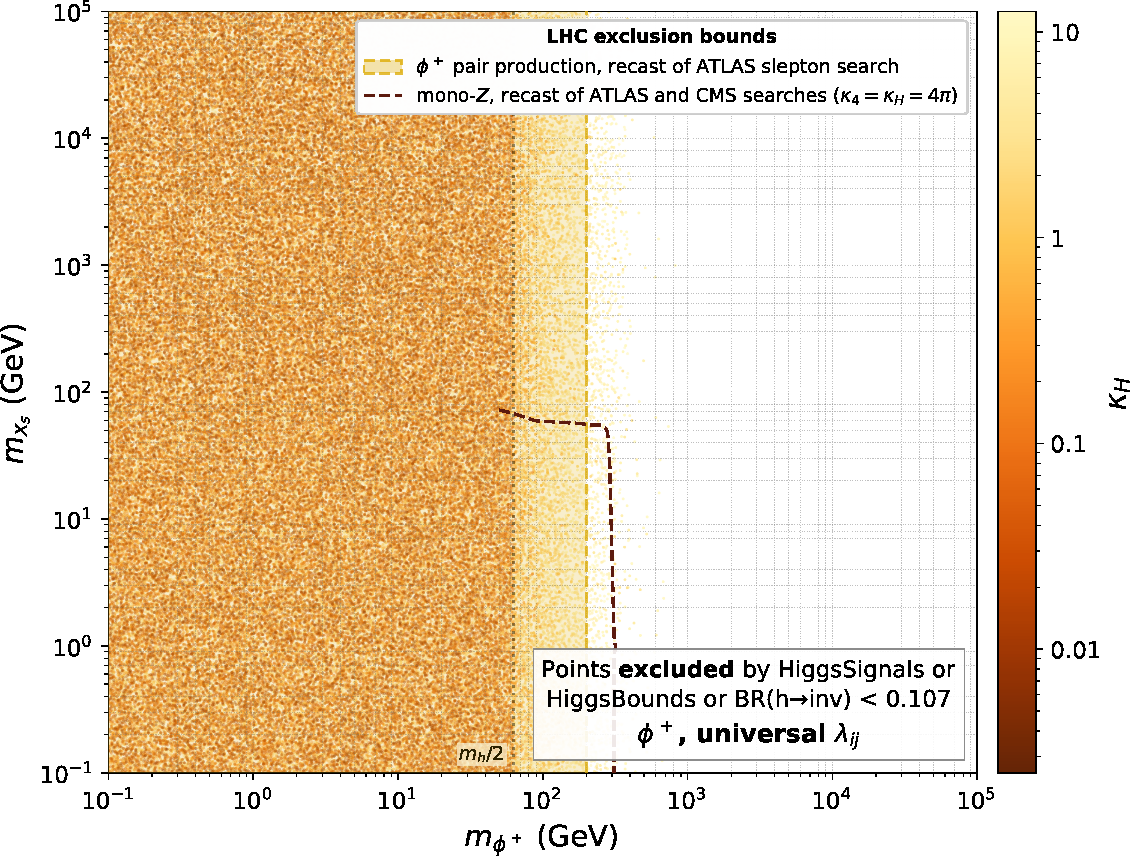,width=.49\textwidth}\hfill
\epsfig{file=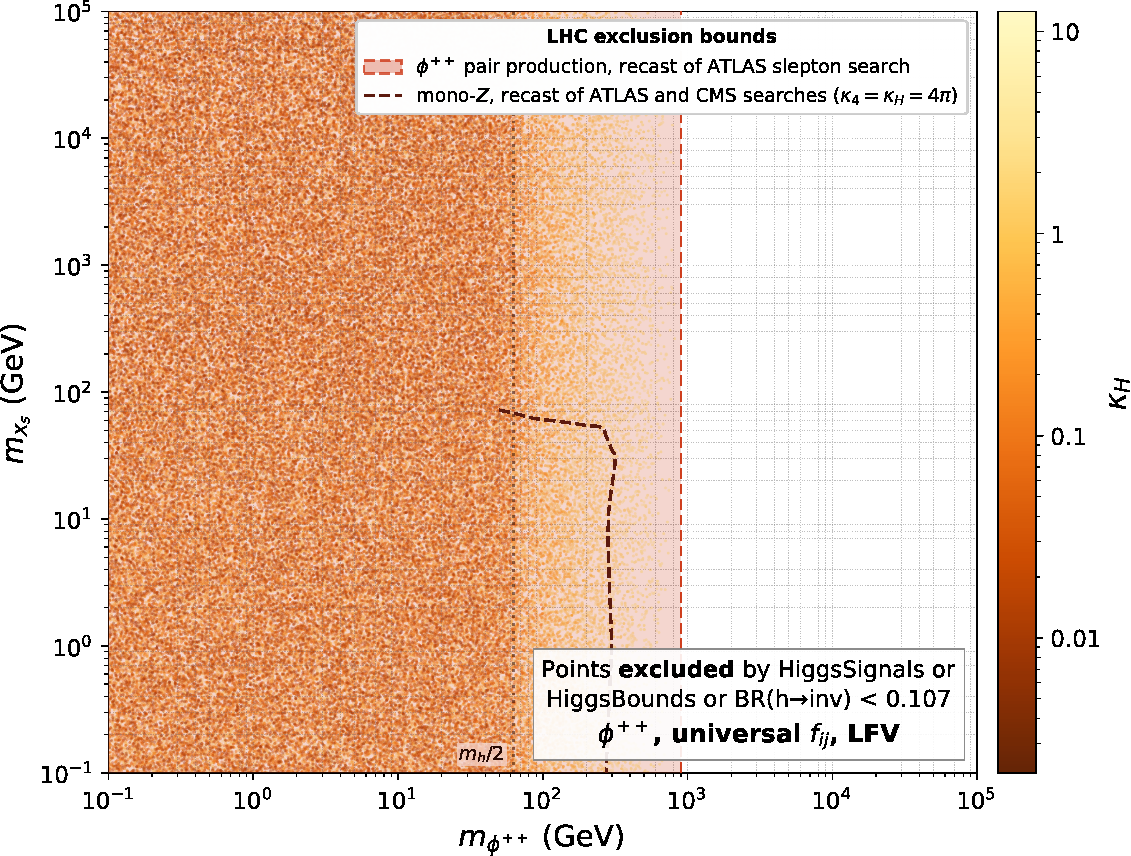,width=.49\textwidth}
\caption{\label{fig:LHCHTBRinv} Points excluded by LHC bounds from pair production of charged scalars and mono-$Z$ processes, and by Higgs observables for the $\phi^+$ (left) and $\phi^{++}$ (right) scenarios with universal couplings, including LFV couplings for $\phi^{++}$.}
\end{figure}

The combination of LHC and Higgs constraints is shown in \Cref{fig:LHCHTBRinv}. The Higgs constraints have the largest impact in the region of parameter space where $m_\phi\lesssim m_h$, but exclusions due to loop and off-shell processes extend to higher masses. In both cases, values of $\kappa_H\lesssim 2\times 10^{-3}$ do not lead to any exclusion. However, the LHC bounds from pair production of the charged scalars are strongly competitive. In the case of $\phi^+$, the mono-$Z$ constraint can even be dominant for $m_{X_s}\lesssim m_h/2$ and $m_\phi\lesssim300$ GeV, but this bound corresponds to $\kappa_4=\kappa_H=4\pi$ and disappears quickly for smaller values of the couplings. Only a few points of the scan are excluded above the LHC mass bounds because of large $h\to\gamma\gamma$ when $\kappa_H\gtrsim 10$ and $m_\phi$ at most around 1 TeV.

\subsubsection{Perspectives at future colliders}
\label{sec:futurecolliders}
%%%%%%%%%%%%%%%%%%%%%%%%%%%%%%%%%%%%%%%%%%%%%%%%%%%%%%%%%%%%%%%%%%%%%%%%%%%%%%

Dedicated studies of $SU(2)_L$-singlet charged scalars at future colliders are limited and are mostly associated to neutrino-mass models. For $\phi^{++}$, the ILC and CLIC can probe the charged-scalar couplings through pair production, single production and off-shell exchange in dilepton final states~\cite{Crivellin:2018ahj,CLIC:2018fvx}; see also Refs.~\cite{Nomura:2017abh,BhupalDev:2018vpr} for complementary studies at future lepton colliders. 
For $\phi^+$, a $1$~TeV ILC with $1.5~\mathrm{ab}^{-1}$ could exclude $m_\phi\lesssim470$~GeV for a $100\%$ $\mu e$ branching ratio, with the reach reduced to about $410$ ($300$)~GeV for $50\%$ ($30\%$) branching fractions~\cite{Yue:2024bsk}. Future $e^+e^-$ colliders can also probe the $x$-channel through mono-photon production~\cite{Crivellin:2020klg}. At a $\sqrt{s}=10$~TeV muon collider with $10~\mathrm{ab}^{-1}$, pair production of singly- and doubly-charged scalars can probe Zee-Babu benchmarks up to $\mathcal{O}(3\text{--}4)$~TeV~\cite{Jueid:2023qcf}. 
At hadron colliders, FCC-hh projections at $\sqrt{s}=100$~TeV reach about $5.4$~TeV ($1~\mathrm{ab}^{-1}$) and $7$~TeV ($3~\mathrm{ab}^{-1}$) for an $SU(2)_L$-triplet doubly-charged scalar decaying to leptons~\cite{Guedes:2601.00083}; the corresponding reach for the singlet $\phi^{++}$ is expected to be smaller because of its lower Drell-Yan production rate.
These projections can probe part of the mediator-mass range left by the present constraints, although the precise reach depends on the flavour structure of the mediator couplings.

\subsection{Flavour constraints}
\label{sec:lfv}
Flavour constraints on the model only concern the charged scalars, since $X_s$ has no direct coupling to SM fermions. The leptonic couplings of $\phi^+$ and $\phi^{++}$ are tightly constrained by low-energy flavour observables. Both cases have been extensively studied in the literature, so the available relations are used directly here, updating only the experimental bounds that have improved since they were derived.
For $\phi^+$, the dominant constraints come from the radiative decays $\ell\to\ell^\prime\gamma$, from $\ell\to\ell^\prime\ell^{\prime(\prime)}\ell^{\prime(\prime)}$, and from $\mu$--$e$ conversion in nuclei, all computed at one loop in~\cite{Crivellin:2020klg}. Only $\mathrm{Br}[\mu\to e\gamma]$ has improved:
\begin{equation}
\mathrm{Br}[\mu\to e\gamma] < 1.5\times10^{-13}~(90\%~\text{C.L.})~\text{\cite{MEGII:2025gzr}}\,.
\label{eq:megii}
\end{equation}

The doubly-charged $\phi^{++}$ generates a richer set of observables, since both LFV and LFC couplings are allowed: the radiative decay $\mu\to e\gamma$, the tree-level (RGE-improved) decays $\ell_a\to 3\ell_{b,c}$, $\mu$--$e$ conversion and muonium-antimuonium oscillation are computed in~\cite{Crivellin:2018ahj}. As for $\phi^+$, $\mathrm{Br}[\mu\to e\gamma]$ is bounded by \cref{eq:megii}, and $\mathrm{Br}[\tau\to 3\mu]$ has improved:
\begin{equation}
\mathrm{Br}[\tau\to 3\mu] < 1.9\times10^{-8}~(90\%~\text{C.L.})~\text{\cite{Belle-II:2024sce}}\,.
\end{equation}

Given that these bounds do not affect production cross-sections but only the width of $\phi$ and its branching ratios, all of these constraints have been applied to filter the scanned points when all bounds are combined in \cref{sec:combinedbounds}.

\subsection{Cosmological and astrophysical bounds} 
 
The $x$-channel model has been implemented in {\sc micrOMEGAs}\cite{Belanger:2001fz,Belanger:2008sj,Belanger:2018ccd,Belanger:2020gnr}, to obtain the constraints from cosmology and astro-particle observations. 
 
\subsubsection{Relic density}
\label{sec:relic-loops}
 
\begin{figure}[h!]
\centering
\begin{picture}(110,60)(0,0)
\Line[dash](10,50)(30,30)
\Text(8,49)[rc]{$X_s$}
\Line[dash](30,30)(10,10)
\Text(8,9)[rc]{$X_s$}
\Line[dash](30,30)(50,45)
\Text(37,40)[cb]{$\phi^-$}
\Line[dash](30,30)(50,15)
\Text(37,19)[ct]{$\phi^+$}
\Line[](50,45)(70,53)
\Text(72,53)[lc]{$\ell^-_i$}
\Line[](70,38)(50,45)
\Text(72,37)[lc]{$\bar \nu_j$}
\Line[](70,23)(50,15)
\Text(72,23)[lc]{$\ell^+_k$}
\Line[](50,15)(70,8)
\Text(72,7)[lc]{$\nu_l$}
\end{picture}
~
\hskip 20pt
~
\begin{picture}(110,60)(0,0)
\Line[dash](10,50)(30,30)
\Text(8,49)[rc]{$X_s$}
\Line[dash](30,30)(10,10)
\Text(8,9)[rc]{$X_s$}
\Line[dash](30,30)(60,45)
\Text(42,40)[cb]{$\phi^{--}$}
\Line[dash](30,30)(60,15)
\Text(42,19)[ct]{$\phi^{++}$}
\Line[](60,45)(80,53)
\Text(82,53)[lc]{$\ell^-_i$}
\Line[](60,45)(80,38)
\Text(82,37)[lc]{$\ell^-_j$}
\Line[](60,15)(80,23)
\Text(82,23)[lc]{$\ell^+_k$}
\Line[](60,15)(80,8)
\Text(82,7)[lc]{$\ell^+_l$}
\end{picture}
\caption{\label{fig:DMann24} DM annihilation into four SM particles: two charged leptons and two neutrinos (left) or four charged leptons (right), for interactions with $\phi^+$ and $\phi^{++}$ respectively.}
\end{figure}
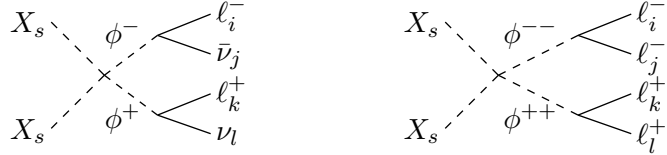

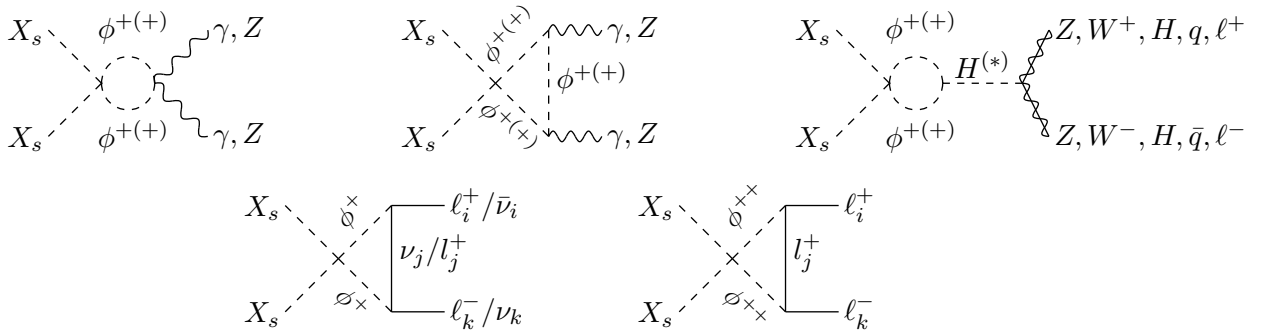
\begin{figure}[h!]
\centering
\begin{picture}(110,60)(0,0)
\Line[dash](10,50)(30,30)
\Text(8,49)[rc]{$X_s$}
\Line[dash](30,30)(10,10)
\Text(8,9)[rc]{$X_s$}
\Arc[dash,clock](40,30)(10,180,0)
\Text(42,45)[cb]{$\phi^{+(+)}$}
\Arc[dash,clock](40,30)(10,0,180)
\Text(42,16)[ct]{$\phi^{+(+)}$}
\Photon(50,30)(70,50){2}{3}
\Text(72,50)[lc]{$\gamma,Z$}
\Photon(50,30)(70,10){2}{3}
\Text(72,10)[lc]{$\gamma,Z$}
\end{picture}
~
\hskip 20pt
~
\begin{picture}(110,60)(0,0)
\Line[dash](10,50)(30,30)
\Text(8,49)[rc]{$X_s$}
\Line[dash](30,30)(10,10)
\Text(8,9)[rc]{$X_s$}
\Line[dash](30,30)(50,50)
\Text(35,36)[cb]{\small\rotatebox{45}{$\phi^{+(+)}$}}
\Line[dash](50,50)(50,10)
\Text(53,32)[lc]{$\phi^{+(+)}$}
\Line[dash](50,10)(30,30)
\Text(35,26)[ct]{\small\rotatebox{-45}{$\phi^{+(+)}$}}
\Photon(50,50)(70,50){2}{3}
\Text(72,50)[lc]{$\gamma,Z$}
\Photon(50,10)(70,10){2}{3}
\Text(72,10)[lc]{$\gamma,Z$}
\end{picture}
~
\hskip 20pt
~
\begin{picture}(140,60)(0,0)
\Line[dash](10,50)(30,30)
\Text(8,49)[rc]{$X_s$}
\Line[dash](30,30)(10,10)
\Text(8,9)[rc]{$X_s$}
\Arc[dash,clock](40,30)(10,180,0)
\Text(42,45)[cb]{$\phi^{+(+)}$}
\Arc[dash,clock](40,30)(10,0,180)
\Text(42,16)[ct]{$\phi^{+(+)}$}
\Line[dash](50,30)(80,30)
\Text(65,32)[cb]{$H^{(*)}$}
\Photon(80,30)(90,50){2}{3}
\Line[](80,30)(90,50)
\Text(92,50)[lc]{$Z,W^+,H,q,\ell^+$}
\Photon(80,30)(90,10){2}{3}
\Line[](80,30)(90,10)
\Text(92,10)[lc]{$Z,W^-,H,\bar q,\ell^-$}
\end{picture}
 
\begin{picture}(110,65)(0,0)
\Line[dash](10,50)(30,30)
\Text(8,49)[rc]{$X_s$}
\Line[dash](30,30)(10,10)
\Text(8,9)[rc]{$X_s$}
\Line[dash](30,30)(50,50)
\Text(35,40)[cb]{\small\rotatebox{45}{$\phi^+$}}
\Line[](50,50)(50,10)
\Text(53,32)[lc]{$\nu_j/l^+_j$}
\Line[dash](50,10)(30,30)
\Text(35,21)[ct]{\small\rotatebox{-45}{$\phi^+$}}
\Line[](70,50)(50,50)
\Text(72,50)[lc]{$\ell^+_i/\bar\nu_i$}
\Line[](50,10)(70,10)
\Text(72,10)[lc]{$\ell^-_k/\nu_k$}
\end{picture}
~
\hskip 20pt
~
\begin{picture}(110,65)(0,0)
\Line[dash](10,50)(30,30)
\Text(8,49)[rc]{$X_s$}
\Line[dash](30,30)(10,10)
\Text(8,9)[rc]{$X_s$}
\Line[dash](30,30)(50,50)
\Text(35,40)[cb]{\small\rotatebox{45}{$\phi^{++}$}}
\Line[](50,50)(50,10)
\Text(53,32)[lc]{$l^+_j$}
\Line[dash](50,10)(30,30)
\Text(35,21)[ct]{\small\rotatebox{-45}{$\phi^{++}$}}
\Line[](50,50)(70,50)
\Text(72,50)[lc]{$\ell^+_i$}
\Line[](50,10)(70,10)
\Text(72,10)[lc]{$\ell^-_k$}
\end{picture}
 
\caption{\label{fig:DMannLoop} DM annihilation into two SM particles via $\phi^+/\phi^{++}$ loops.}
\end{figure}

The topologies through which $X_s$ can annihilate into SM particles can be either at tree- or loop-level, shown in \Cref{fig:DMann24,fig:DMannLoop} respectively, and their relative contribution depends on the ratio between the DM and mediator masses. If $m_{X_s}\gtrsim m_\phi$ the charged scalars can be produced on shell and decay into leptons, while in the opposite regime the loop topologies become increasingly dominant as the $m_{X_s}/m_\phi$ ratio decreases. The dominant annihilation channels considered here carry an overall $\kappa_4^2$ dependence, with additional $\kappa_H$ dependence entering the Higgs-mediated channels. 
 
The tree-level channels of \cref{fig:DMann24} are computed with {\sc micrOMEGAs}, including the off-shell $2\to3$ and $2\to4$ processes near the threshold $m_{X_s}\simeq m_\phi$.
The loop-induced $2\to2$ channels of \cref{fig:DMannLoop} are evaluated independently (see \cref{app:relicloops}) and passed to {\sc micrOMEGAs} through the \texttt{improveCrossSection} function.
 
\begin{figure}[h]
\centering
\epsfig{file=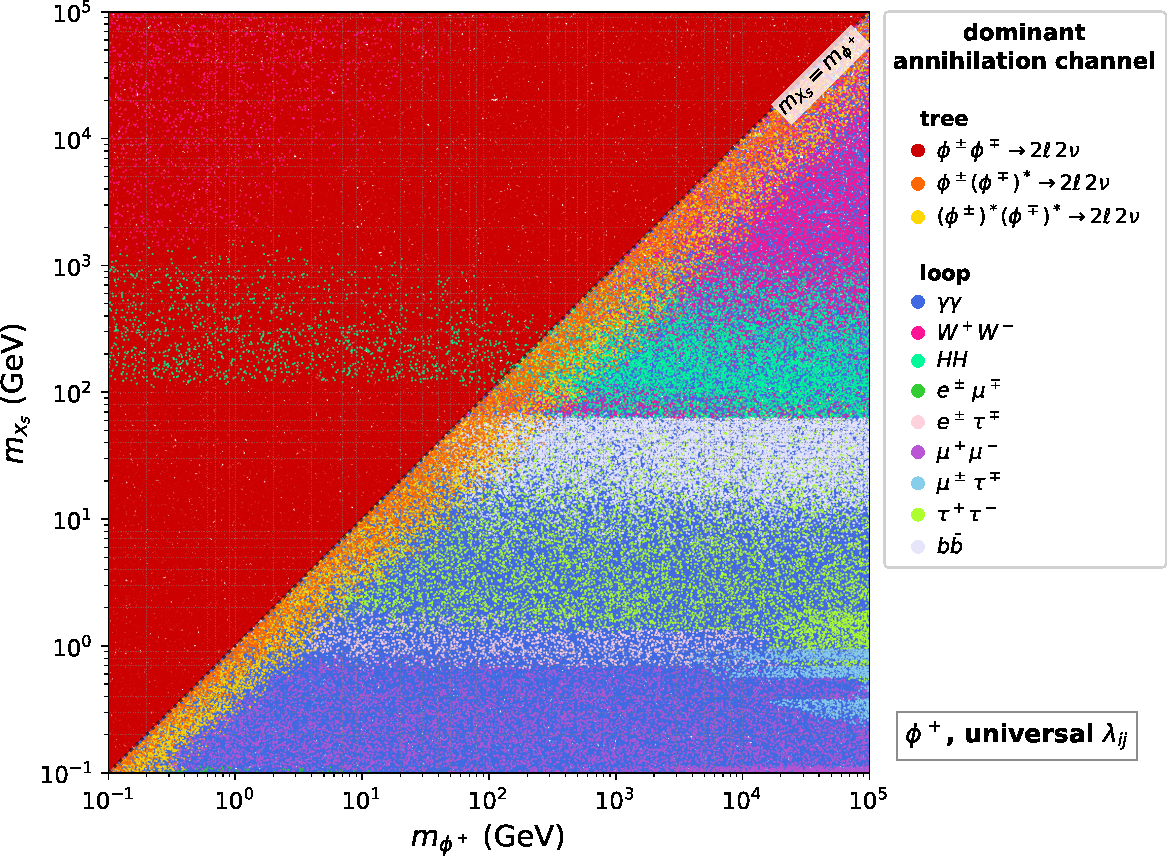,width=.49\textwidth}\hfill
\epsfig{file=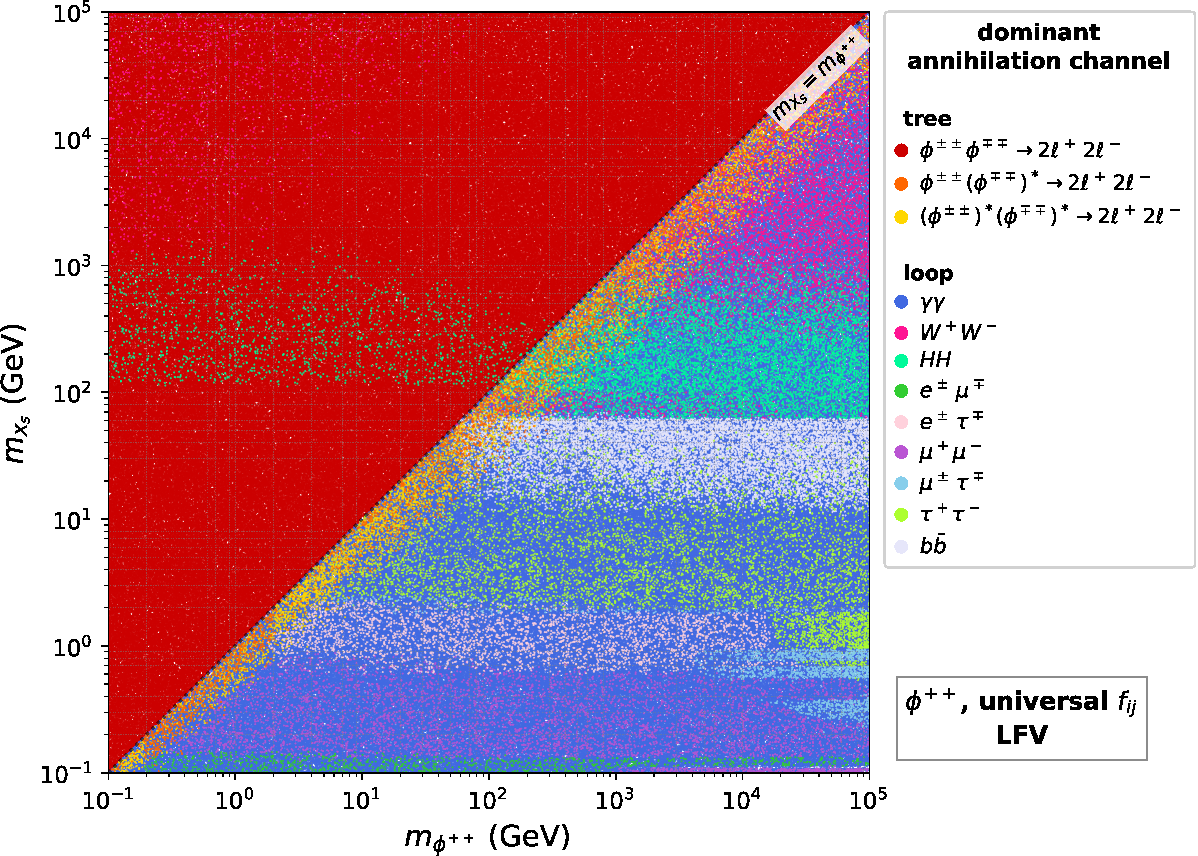,width=.49\textwidth}\\
\epsfig{file=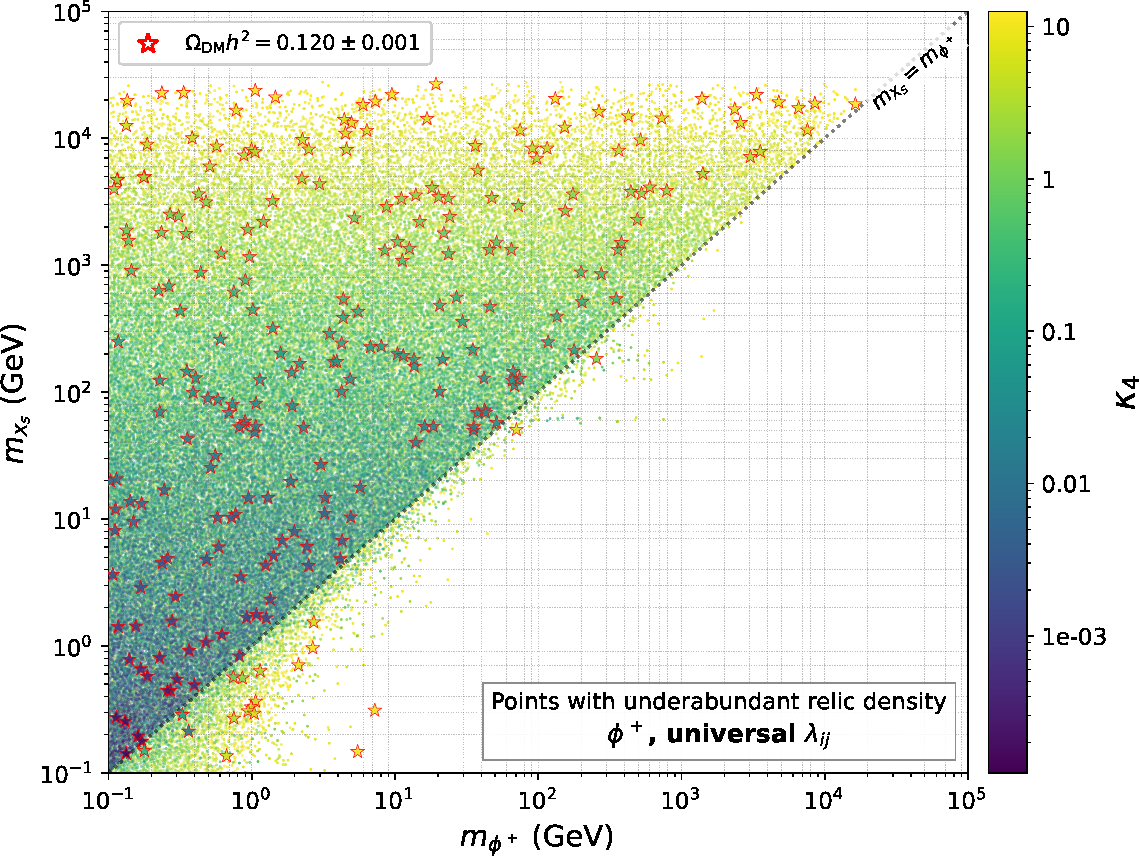,width=.49\textwidth}\hfill
\epsfig{file=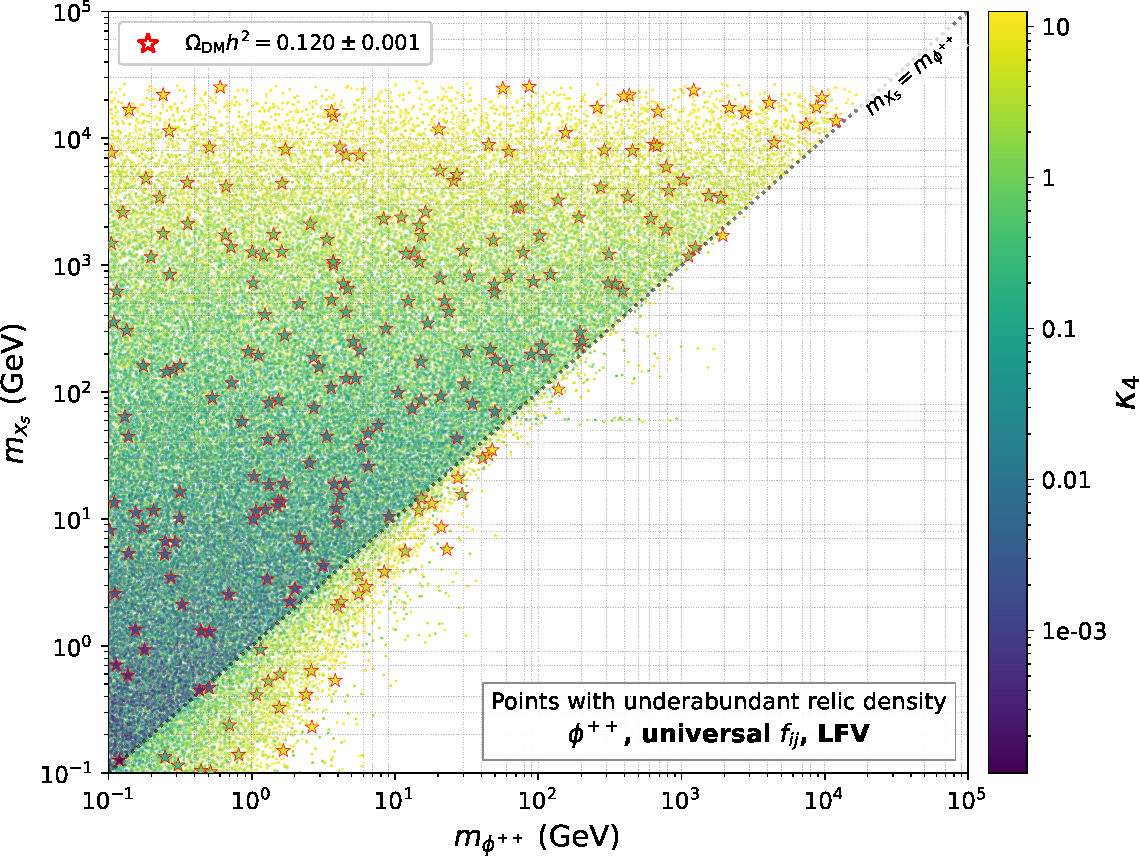,width=.49\textwidth}
\caption{\label{fig:relics1universal} {\bf Top panels:} The dominant annihilation process, i.e.\ the channel with the largest contribution to the total thermally averaged annihilation rate, for each point of the scan. {\bf Bottom panels:} points with $\Omega h^2\leq0.121$, where the colour of the points is associated to the value of $\kappa_4$ and stars identify points consistent with the measured relic density within 1$\sigma$. In both rows the results are for $\phi^+$ {\bf (left)} and $\phi^{++}$  {\bf(right)}.}
\end{figure}

The top panels of \cref{fig:relics1universal} show the dominant annihilation process as a function of the DM and $\phi$ mass for both $\phi^+$ and $\phi^{++}$. For $m_\phi<m_{X_s}$, tree-level annihilation into $\phi$ pairs through the four-leg $\kappa_4$ vertex dominates. Near the kinematic threshold $m_\phi\simeq m_{X_s}$, the $2\to3$ and $2\to4$ final states take over, and for $m_\phi>m_{X_s}$ annihilation to photons dominates almost everywhere, with other channels becoming relevant only near specific mass thresholds. The $HH$ final state can compete with the tree-level channels even for $m_\phi>m_{X_s}$ when $\kappa_H$ is large.

The bottom panels of \cref{fig:relics1universal} show which points are underabundant or at most $1\sigma$ above the {\sc Planck} measurement~\cite{Planck:2018vyg},
\begin{equation}
\Omega^{\text{Planck}}_{\text{DM}}h^2 = 0.120 \pm 0.001 \;,
\label{eq:planck}
\end{equation} 
This condition leaves points predominantly in the region $m_{X_s}\gtrsim m_\phi$. The point are further identified by their value of $\kappa_4$ and marked as stars when they reconstruct the correct relic density within $1\sigma$. Small couplings are sufficient for the required annihilation at low $m_{X_s}$ and $m_\phi$. As the DM mass increases, the minimum $\kappa_4$ needed for efficient annihilation grows, eventually reaching values for which the perturbative treatment is no longer reliable. A few points survive with $m_{X_s}<m_\phi$, mostly at small $m_\phi$ (where the loop suppression is mild, but already excluded by the LHC bounds of \cref{sec:LHC}) or in a narrow window around $m_\phi\sim100$~GeV where the $WW$ and $HH$ channels become relevant alongside the off-shell tree-level processes. 
Points giving the measured relic density are evenly spread and do not show specific patterns in the parameter space.

\subsubsection{Direct detection}
\label{sec:dd}
 
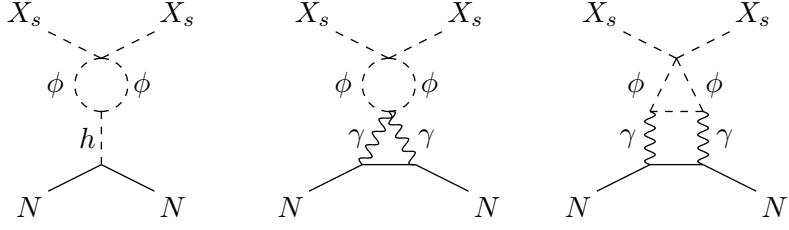
\begin{figure}[h]
\centering
\begin{picture}(60,80)(0,0)
\Line[dash](10,70)(30,60)
\Text(8,72)[br]{$X_s$}
\Line[dash](30,60)(50,70)
\Text(52,72)[bl]{$X_s$}
\Arc[dash,clock](30,50)(10,90,270)
\Text(16,50)[rc]{$\phi$}
\Arc[dash,clock](30,50)(10,270,90)
\Text(42,50)[lc]{$\phi$}
\Line[dash](30,20)(30,40)
\Text(28,30)[rc]{$h$}
\Line[](10,10)(30,20)
\Text(8,8)[tr]{$N$}
\Line[](30,20)(50,10)
\Text(52,8)[tl]{$N$}
\end{picture}
~
\hskip 20pt
~
\begin{picture}(70,80)(0,0)
\Line[dash](20,70)(40,60)
\Text(18,72)[br]{$X_s$}
\Line[dash](40,60)(60,70)
\Text(62,72)[bl]{$X_s$}
\Arc[dash,clock](40,50)(10,90,270)
\Text(26,50)[rc]{$\phi$}
\Arc[dash,clock](40,50)(10,270,90)
\Text(52,50)[lc]{$\phi$}
\Photon(30,20)(40,40){2}{4}
\Text(31,30)[cr]{$\gamma$}
\Photon(40,40)(50,20){2}{4}
\Text(51,30)[cl]{$\gamma$}
\Line[](10,10)(30,20)
\Text(8,8)[tr]{$N$}
\Line[](30,20)(50,20)
\Line[](50,20)(70,10)
\Text(72,8)[tl]{$N$}
\end{picture}
~
\hskip 20pt
~
\begin{picture}(70,80)(0,0)
\Line[dash](20,70)(40,60)
\Text(18,72)[br]{$X_s$}
\Line[dash](40,60)(60,70)
\Text(62,72)[bl]{$X_s$}
\Line[dash](30,40)(40,60)
\Text(28,50)[rc]{$\phi$}
\Line[dash](50,40)(40,60)
\Text(51,50)[lc]{$\phi$}
\Line[dash](30,40)(50,40)
\Photon(30,20)(30,40){2}{4}
\Text(25,30)[cr]{$\gamma$}
\Photon(50,40)(50,20){2}{4}
\Text(55,30)[cl]{$\gamma$}
\Line[](10,10)(30,20)
\Text(8,8)[tr]{$N$}
\Line[](30,20)(50,20)
\Line[](50,20)(70,10)
\Text(72,8)[tl]{$N$}
\end{picture}
\caption{\label{fig:DDdiagrams} Topologies for the scattering of $X_s$ on a target nucleus $N$ for direct detection: Higgs exchange {\bf (left)} and two-photon exchange through the bubble {\bf (centre)} and triangle {\bf (right)} topologies.}
\end{figure}

In absence of a direct tree-level coupling with the Higgs boson, $X_s$ scatters against nuclei only through the one- and two-loop topologies of \cref{fig:DDdiagrams}. 

The amplitude of the Higgs diagram is governed by $\lambda_{4,\rm eff}(t;\mu_0)$ at $t=-q^2$, where $q^2=2m_NE_R$ is the squared momentum transfer to a nucleus $N$ of mass $m_N$ recoiling with energy $E_R$. For $\mu_0=m_\phi$, $\lambda_{4,\rm eff}(0;m_\phi)=0$, and for $|t|\ll m_\phi^2$ \cref{eq:fsmallx} gives
\begin{equation}
\label{eq:lam4efft}
\lambda_{4,\rm eff}(t;m_\phi)\simeq-\frac{\kappa_4\kappa_H}{96\pi^2}\,\frac{t}{m_\phi^2}\,,
\end{equation}
leading to the known cross section of the real-scalar Higgs portal~\cite{Burgess:2000yq,Cline:2013gha} with $\lambda_4$ replaced by $\lambda_{4,\rm eff}(t;\mu_0)$,
\begin{equation}
\label{eq:sigmaSIh}
\sigma^{\rm SI}_{n,h}(t)=\frac{\mu_n^2\,m_n^2\,f_n^2}{4\pi\,m_{X_s}^2}\,\frac{|\lambda_{4,\rm eff}(t;\mu_0)|^2}{(t-m_h^2)^2}\,,
\end{equation}
where $m_n$ is the nucleon mass, $\mu_n$ the $X_s$-nucleon reduced mass and $f_n=\sum_{q=u,d,s}f^n_{Tq}+\tfrac29f^n_{TG}$ the nucleon scalar form factor, with $f^n_{TG}=1-\sum_{q=u,d,s}f^n_{Tq}$. 
The direct-detection routine of {\sc micrOMEGAs} is used to compare with data from XENON1T~\cite{XENON:2018voc}, LZ~\cite{LZ:2024zvo}, DarkSide-50, PICO-60 and CRESST-III. Underabundant points have their rate rescaled by $\Omega h^2/0.12$.
Within the range of the scan, this process does not exclude any point, as it is expected by how the $x$-channel mechanism is constructed. \\
 
The two-photon diagrams of \cref{fig:DDdiagrams}, on the other hand, do not depend on the $x$-channel condition and can be described at effective field theory level by the dimension-six Rayleigh operator~\cite{Weiner:2012cb},
\begin{equation}
\label{eq:rayleigh_op}
\mathcal L_{\rm EFT}\supset\tilde{\mathcal C}_{\gamma\gamma}\,X_s^2F_{\mu\nu}F^{\mu\nu}\;.
\end{equation}
Following Ref.~\cite{Barducci:2025twe}, $\tilde{\mathcal C}_{\gamma\gamma}$ is fixed by matching. For $s\ll m_\phi^2$ the scalar $\phi$ is integrated out, and the operator must reproduce the $X_sX_s\to\gamma\gamma$ amplitude of the full model, \cref{eq:MggXsXs}, which for $1+2m_\phi^2C_0\simeq-s/12m_\phi^2$ becomes
\begin{equation}
\mathcal M_{\rm UV}=\frac{Q_\phi^2\,\alpha\kappa_4}{24\pi m_\phi^2}\,s\;\varepsilon_3\!\cdot\!\varepsilon_4\;.
\end{equation}
The operator gives
\begin{equation}
\mathcal M_{\rm EFT}=-8\,\tilde{\mathcal C}_{\gamma\gamma}\big[(k_3\!\cdot\!k_4)(\varepsilon_3\!\cdot\!\varepsilon_4)-(k_3\!\cdot\!\varepsilon_4)(k_4\!\cdot\!\varepsilon_3)\big]=-4\,\tilde{\mathcal C}_{\gamma\gamma}\,s\;\varepsilon_3\!\cdot\!\varepsilon_4\;,
\end{equation}
where the factor 8 accounts for the two $X_s$ fields, the two photons from the field strengths, and a further factor 2 from the antisymmetric $k_\mu\varepsilon_\nu-k_\nu\varepsilon_\mu$ in their contraction, and where for transverse photons in the centre-of-mass frame $k_i\!\cdot\!\varepsilon_j=0$ and $k_3\!\cdot\!k_4=s/2$. Equating the two amplitudes fixes
\begin{equation}
\label{eq:CtildeGG}
\tilde{\mathcal C}_{\gamma\gamma}=-\frac{Q_\phi^2\,\alpha\kappa_4}{96\pi m_\phi^2}\;.
\end{equation}
Scattering against the target nucleus $N$ gives:
\begin{equation}
\frac{d\sigma^{\rm Ray}}{dq^2}\simeq\frac{2\alpha^2Q_0^2}{\pi^2m_{X_s}^2v_{\rm rel}^2}\,\tilde{\mathcal C}_{\gamma\gamma}^2\,Z_N^4\,F^2_{\rm Ray}(q/Q_0)\;,
\end{equation}
where $Z_N$ and $A$ are the charge and mass number of $N$, $v_{\rm rel}$ is the relative velocity, $Q_0=0.48/(0.3+0.89A^{1/3})$~GeV is the nuclear coherence scale and $F_{\rm Ray}$ is the nuclear two-photon response, normalised to $F_{\rm Ray}(0)=1$~\cite{Frandsen:2012rk}.
Since $F_{\rm Ray}$ depends only on $q/Q_0$, the cross section integrated over the recoil spectrum scales as
\begin{equation}
\label{eq:sigmaRay0}
\sigma^{\rm Ray}\;\propto\;\tilde{\mathcal C}_{\gamma\gamma}^2\;\propto\;\frac{Q_\phi^4\kappa_4^2}{m_\phi^4}\;.
\end{equation}
 
The Rayleigh cross section has been interfaced in {\sc micrOMEGAs}. The contact operator keeps only the leading term of the loop function of \cref{eq:MggXsXs}, $1+2m_\phi^2C_0\simeq-(s/12m_\phi^2)\,[1+2s/15m_\phi^2+\dots]$, where $s$ is replaced by $t=-q^2$ in scattering. At the largest recoil energy implemented in {\sc micrOMEGAs}, $E_R=80$~keV, which gives $q^2\simeq0.02$~GeV$^2$ in Xenon, the contribution of the first neglected term grows from $1\%$ at $m_\phi=0.51$~GeV to $26\%$ at $m_\phi=0.1$~GeV, so the validity of the contact operator is questionable at the lower end of this range.

\begin{figure}[h]
\centering
\epsfig{file=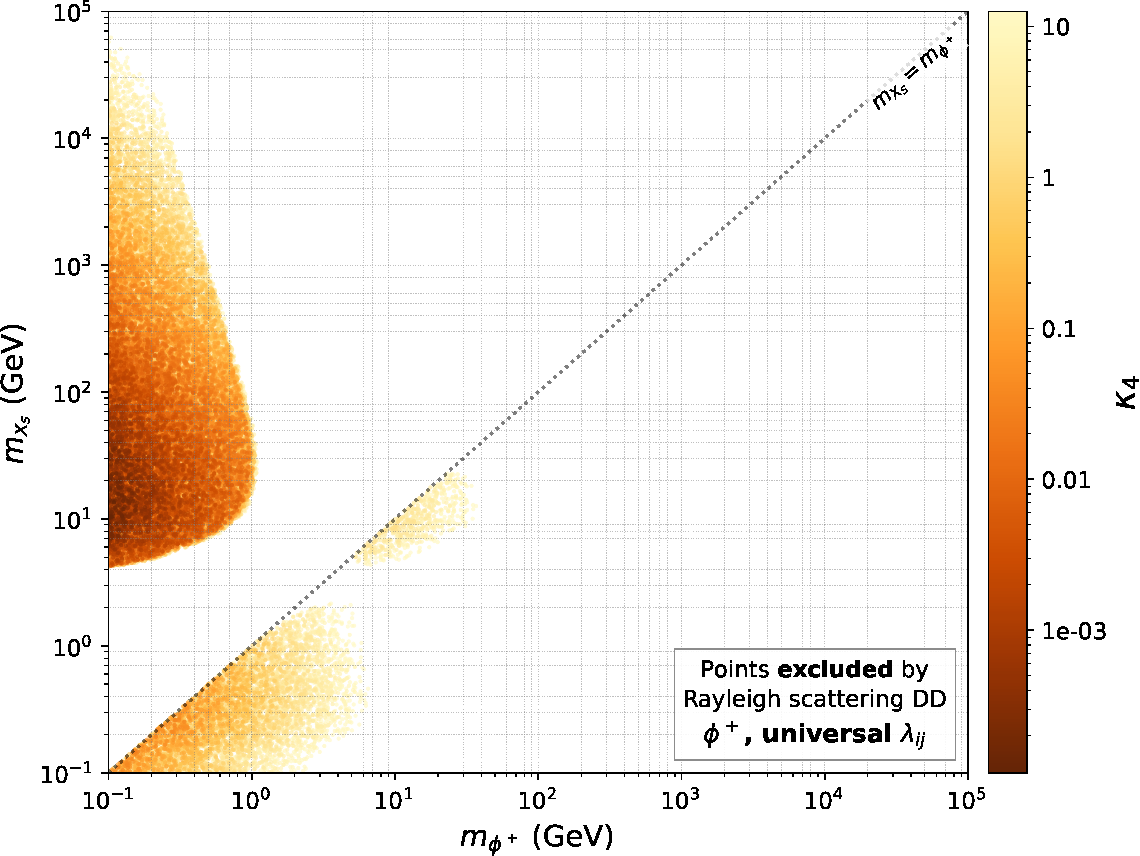,width=.49\textwidth}
\epsfig{file=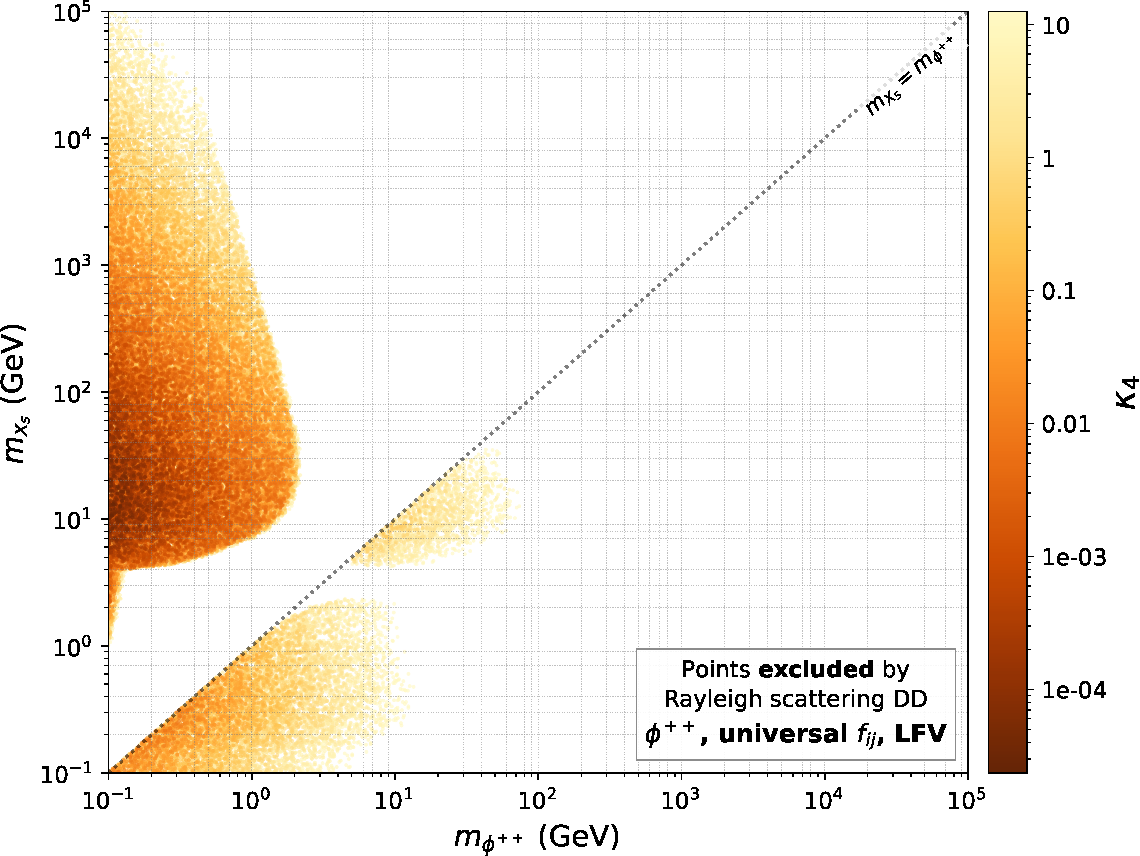,width=.49\textwidth}
\caption{\label{fig:excludedDDs1universal} Points excluded by direct detection via Rayleigh scattering, where the colour gradient represents the value of $\kappa_4$, for $\phi^+$ {\bf(left)} and $\phi^{++}$ {\bf(right)}.}
\end{figure}

\cref{fig:excludedDDs1universal} displays the points excluded by direct detection via Rayleigh scattering in the plane of $m_\phi$ and $m_{X_s}$, for $\phi^+$ (left) and $\phi^{++}$ (right), coloured according to the value of $\kappa_4$. For $m_{X_s}>m_\phi$ the rescaling by $\Omega h^2/0.12$ cancels the $\kappa_4^2$ of \cref{eq:sigmaRay0} above a value $\kappa_4$ which leads to underabundance, confining the excluded region to $m_\phi\lesssim1$~GeV and producing the drop-like shape of the region. For $m_{X_s}<m_\phi$ no rescaling applies and the reach in $m_\phi$ grows with $\kappa_4$. This gives the two regions along the diagonal, excluded by CRESST at the lowest masses and XENON1T above a few GeV.

\subsubsection{Indirect detection}
\label{sec:id}

For indirect detection, the same channels entering the relic density (\cref{sec:relic-loops}) must be considered with present-day cross section $\langle\sigma v_{\rm rel}\rangle_0$ at the Galactic relative velocity $v_{\rm rel}\simeq10^{-3}$, rescaled by a factor $(\Omega h^2/0.120)^2$ for underabundant points.

Different indirect observables are considered: CMB energy-injection constraints from {\sc Planck}~\cite{Planck:2018vyg,Slatyer:2015jla,Leane:2018kjk}, which restrict the amount of energy deposited during recombination that would otherwise alter the CMB ionization history and anisotropy spectrum; Fermi-LAT dwarf spheroidal galaxy observations~\cite{Fermi-LAT:2015att} (for $m_{X_s}\gtrsim0.9$~GeV); and galactic center gamma-ray line searches from Fermi-LAT~\cite{Fermi-LAT:2015kyq} and H.E.S.S.~\cite{HESS:2018kom}.

While the first is evaluated using a native {\sc micrOMEGAs} function, the one-loop annihilation cross-sections for the monochromatic $\gamma\gamma$ and $\gamma Z$ lines and the continuum photon spectra are calculated using {\sc micrOMEGAs} and subsequently compared against experimental bounds using data implemented in {\sc MadDM}~\cite{Backovic:2013dpa,Backovic:2015cra,Ambrogi:2018jqj,Arina:2021gfn}.

The photon signal depends on the ratio between $m_{X_s}$ and $m_\phi$. For $m_{X_s}<m_\phi$ the $X_sX_s\to\phi\phi^\dagger$ channel is closed, and the loop-induced $\gamma\gamma$ and $\gamma Z$ lines carry most of $\langle\sigma v_{\rm rel}\rangle_0$. For $m_{X_s}>m_\phi$ the tree-level channel $X_sX_s\to\phi\phi^\dagger\to2\ell2\nu$ ($4\ell$ for $\phi^{++}$) dominates the annihilation and yields photons only from radiation and $\tau$ decays; for $\phi^+$ the lines then amount to a tiny fraction of the total rate.

\begin{figure}[h]
\centering
\epsfig{file=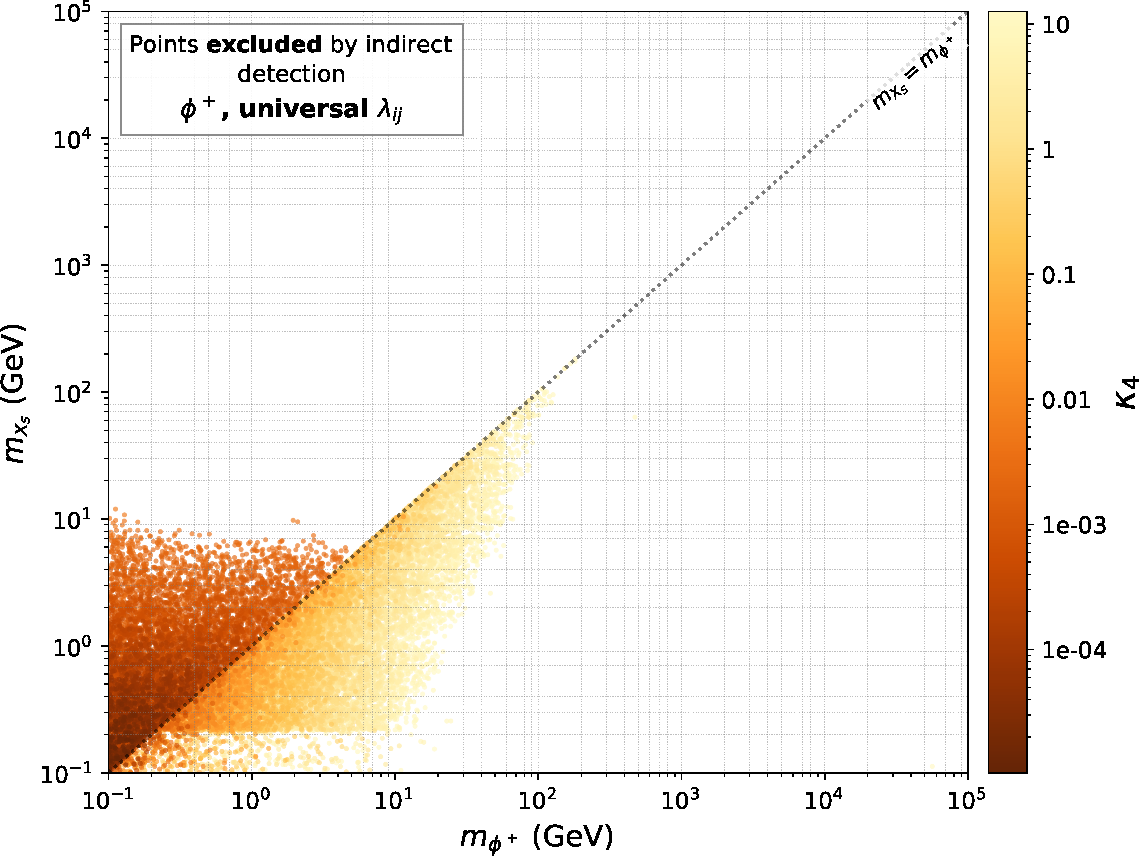,width=.49\textwidth}
\epsfig{file=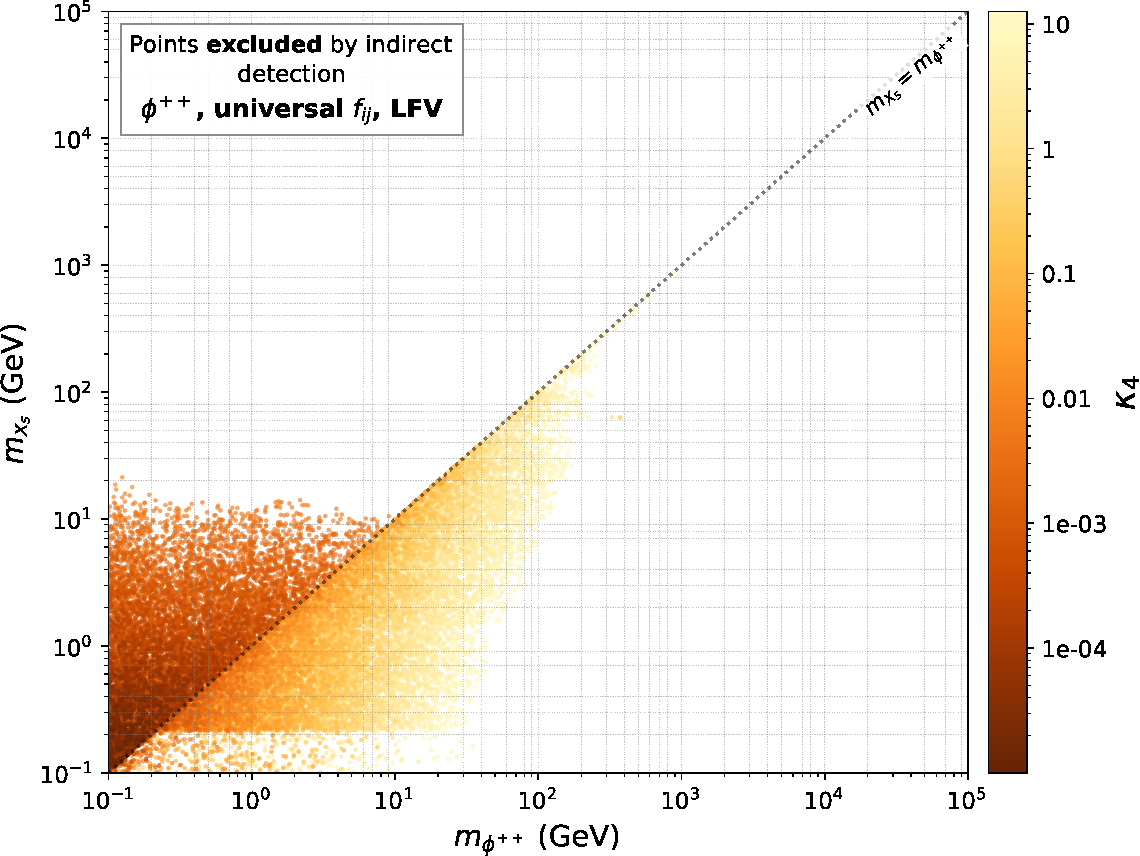,width=.49\textwidth}
\caption{\label{fig:excludedIDs1universal} Points excluded by indirect detection, coloured by $\kappa_4$, for $\phi^+$ {\bf (left)} and $\phi^{++}$ {\bf (right)} with universal couplings, including LFV couplings for $\phi^{++}$.}
\end{figure}

\Cref{fig:excludedIDs1universal} presents the parameter space excluded at 95\% C.L. by the indirect detection observables. 
Near and below the $m_{X_s}=m_\phi$ diagonal, the exclusion is driven mainly by the Fermi-LAT searches for the monochromatic $\gamma\gamma$ and $\gamma Z$ lines, with the Planck CMB bound contributing mostly at $m_{X_s}\lesssim0.3$~GeV. 
Above the diagonal, only the Planck CMB bound plays a role in excluding points. The exclusion is a band in $\kappa_4$ around the value giving $\Omega h^2=0.12$, and large couplings are not excluded in this region.

The exclusion is stronger for $\phi^{++}$. The $\gamma\gamma$ and $\gamma Z$ rates scale as $Q_\phi^4\kappa_4^2$ (\cref{eq:MggXsXs}), so for same masses and couplings the line signal is $16$ times larger than for $\phi^+$. The excluded region therefore extends to smaller $\kappa_4$ and to larger masses around the $m_{X_s}=m_\phi$ diagonal. The flavour-violating couplings of $\phi^{++}$ further enlarge the region affected by indirect constraints through the continuum channels.

\subsection{Combined bounds}
\label{sec:combinedbounds}

\begin{figure}[h]
\centering
\epsfig{file=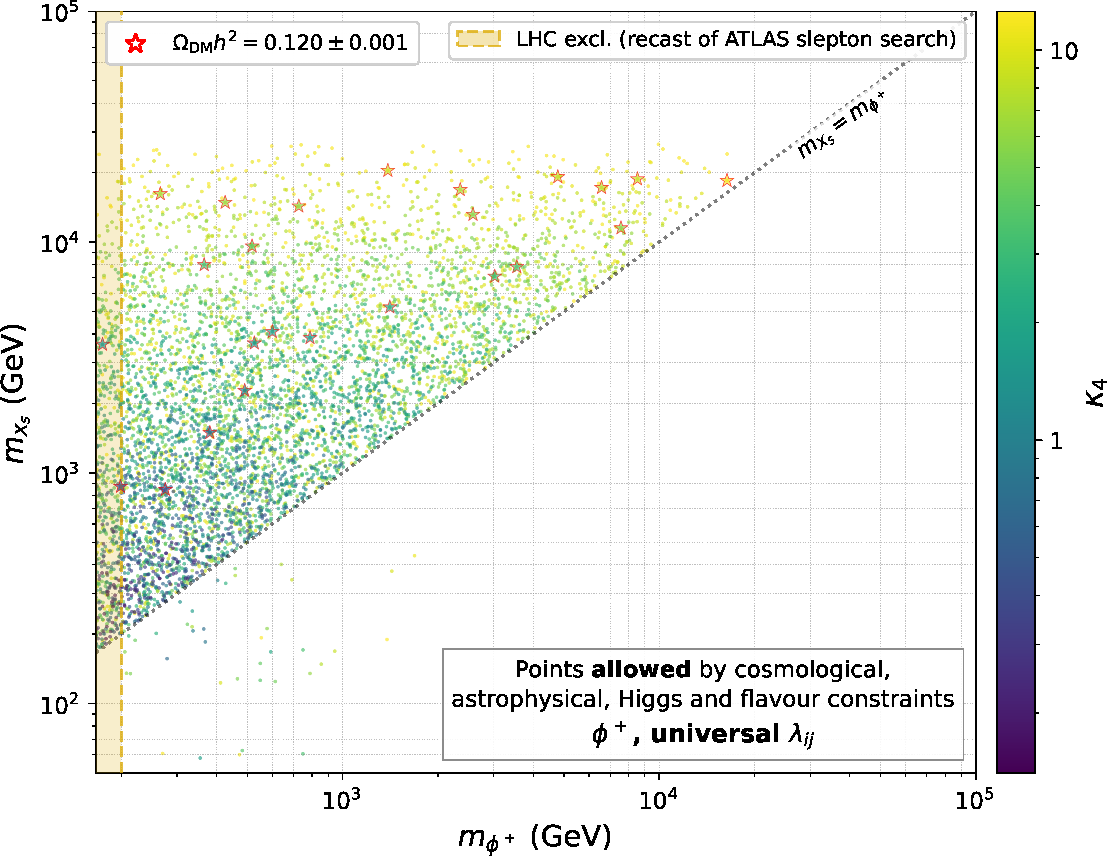,width=.49\textwidth}\hfill
\epsfig{file=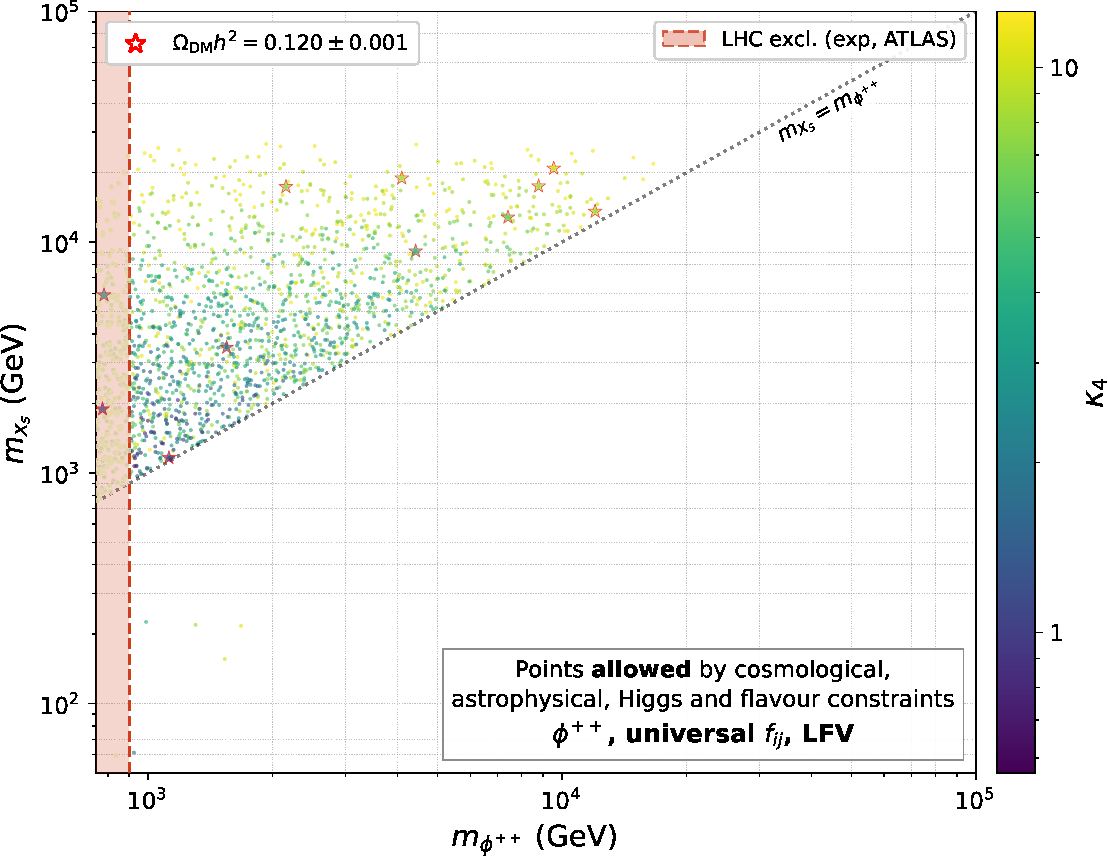,width=.49\textwidth}
\caption{\label{fig:combinedbounds} Points allowed by the Higgs, flavour, relic-density, direct and indirect detection constraints, coloured by $\kappa_4$, for $\phi^+$ {\bf (left)} and $\phi^{++}$ {\bf (right)} with universal couplings. The shaded bands denote the regions excluded by the LHC searches for pair production of the charged scalars.}
\end{figure}

The points which survive the combined Higgs, flavour, relic-density, direct and indirect detection constraints are shown in \cref{fig:combinedbounds}.
For both charges the surviving points overwhelmingly lie at $m_{X_s}\gtrsim m_\phi$, where the relic density is given by the annihilation into on-shell $\phi$ pairs and, close to threshold, by the off-shell $2\to3$ and $2\to4$ processes of \cref{sec:relic-loops}. 
For $m_\phi$ above the collider limits, neither direct or indirect detection are sensitive enough. The surviving parameter space of the minimal $x$-channel scenario with electrically charged scalar mediators can therefore be probed directly only at future colliders.

\section{Possible non-minimal realisations}
\label{sec:nonminimal}

The singly- and doubly-charged singlets studied in the previous sections are one realisation of the $x$-channel mechanism. A different gauge representation for $\phi$ changes its production rate and the SM operators through which it decays, completely changing the phenomenology.

Scalar leptoquarks, for example, couple to quark-lepton bilinears, such as $\overline{Q_L^c}L_L\phi$ or $\overline{u_R^c}e_R\phi$~\cite{Dorsner:2016wpm}; scalar diquarks in colour-triplet or sextet representations couple to same-chirality quark pairs, as $\overline{u_R^c}d_R\phi$ or $\overline{Q_L^c}Q_L\phi$~\cite{Carpenter:2023diquark}; colour-octet electroweak doublets couple to quark-antiquark bilinears~\cite{Manohar:2006ga,Carpenter:2023diquark}. A scalar doublet with hypercharge $Y=3/2$, containing a singly- and a doubly-charged state, has no Yukawa coupling to SM leptons but can still decay via different mechanisms~\cite{Aoki:2011yk}. Furthermore, a renormalisable decay is not a strict requirement. In general, operators of the form $\Lambda^{4-d}\,\phi\,\mathcal O_{\rm SM}$ give decays into fermions, gauge bosons or Higgs bosons, and a large $\Lambda$ can lead to long-lived mediators. If $\phi$ carries a confining gauge charge, bound states containing $\phi$ become the observable particles~\cite{Cline:2012nw}. 

All these cases can preserve the $x$-channel structure and lead to collider, flavour and cosmological phenomenology that differs from the minimal leptonic realisation studied here.

%%%%%%%%%%%%%%%%%%%%%%%%%%%%%%%%%%%%%%%%%%%%%%%%%%%%%%%%%%%%%%%%%%%%%%%%%%%%%%
\section{Conclusion}
\label{sec:conclusion}

I have studied a class of scenarios in which a scalar singlet dark matter candidate $X_s$ communicates with the Standard Model through the quartic coupling $X_s^2\phi^\dagger\phi$ to a scalar mediator $\phi$ carrying a conserved internal quantum number and belonging to a representation with no components that can mix with the Higgs boson, so that the dark matter interacts only with exotic states. The Higgs portal is suppressed by the $x$-channel mechanism, defined by imposing a boundary condition $\lambda_4(\mu_0)=0$ on the running Higgs-portal coupling at $\mu_0=m_\phi$. At one loop this condition removes Higgs-mediated amplitudes at zero momentum transfer while leaving a finite momentum-dependent contribution. The condition does not depend on the internal quantum numbers of the mediator. A minimal realisation with singly- and doubly-charged $SU(2)_L$ singlets has been analysed in detail, finding that direct and indirect detection experiments are not sensitive enough to probe the region of parameter space allowed by underabundant relic density and LHC bounds. Pair production of the charged mediator at future colliders can improve the sensitivity reach on this realisation.

The $x$-channel mechanism applies to complex dark matter scalars and to mediators with fractional charge, with colour, or in non-trivial $SU(2)_L$ representations, opening phenomenological directions which can be analysed in future studies.

\section*{Note added}

While this work was being completed, Ref.~\cite{Borah:2026aqv} appeared, studying real singlet dark matter with a charged singlet scalar and a charged-scalar portal. There is therefore direct overlap at the level of the field content of the minimal $x$-channel realisation and of the quartic interaction connecting the dark matter to the charged mediator. Ref.~\cite{Borah:2026aqv} develops the thermal scalar potential, $U(1)_Y$ non-restoration, first-order phase transitions and gravitational-wave phenomenology not treated in this analysis, whereas the present analysis focus on the $x$-channel mechanism and uses the charged scalar scenario as minimal example, extending to the doubly-charged mediator case, and studying the combined constraints from collider, Higgs, flavour, and cosmological/astrophysical observables.

%%%%%%%%%%%%%%%%%%%%%%%%%%%%%%%%%%%%%%%%%%%%%%%%%%%%%%%%%%%%%%%%%%%%%%%%%%%%%%
% \section*{Acknowledgements}

%%%%%%%%%%%%%%%%%%%%%%%%%%%%%%%%%%%%%%%%%%%%%%%%%%%%%%%%%%%%%%%%%%%%%%%%%%%%%%
\appendix

\section{Renormalisation of the Higgs-portal operator}
\label{app:renorm-derivation}

This appendix derives \cref{eq:lam4eff}. The bubble diagram of \cref{fig:H2X2operator} is UV divergent and contributes to the renormalisation of $\lambda_4$. Its expression is given in terms of the Passarino-Veltman (PV) function $B_0$~\cite{Passarino:1978jh} as 
\begin{equation}
\label{eq:hhXsbubble}
i\Gamma^{(4)}_{\rm bub} = \frac{i\kappa_4\kappa_H}{16\pi^2} B_0(q^2;m_\phi^2,m_\phi^2)\,.
\end{equation}

In $d=4-2\epsilon$ dimensions the bare coupling of the tree-level contact term is $\lambda_4^{\rm bare}=\mu^{2\epsilon}\left[\lambda_4(\mu)+\delta\lambda_4\right]$, and the same factor $\mu^{2\epsilon}$ accompanies each quartic coupling in the loop. At leading order in $\epsilon$ the full $(H^\dagger H) X_s^2$ vertex is then
\begin{equation}
\label{eq:lam4effbeforect}
\lambda_{4,\rm eff}(q^2;\mu)
=
\lambda_4(\mu)+\delta\lambda_4
-\frac{\kappa_4\kappa_H}{16\pi^2}
B_0(q^2;m_\phi^2,m_\phi^2;\mu)\,.
\end{equation}
The PV function is expressed as usual as a sum of a divergent and of a finite term as
\begin{equation}
\label{eq:B0decomp}
B_0(q^2;m_\phi^2,m_\phi^2;\mu)
=
\Delta
+
B_0^{\rm fin}(q^2;m_\phi^2,m_\phi^2;\mu)
+\mathcal O(\epsilon)\,,\qquad
\Delta
\equiv
\frac{1}{\epsilon}
-\gamma_E
+\ln 4\pi\,.
\end{equation}
Inserting \cref{eq:B0decomp} in \cref{eq:lam4effbeforect} gives
\begin{equation}
\label{eq:lam4effpole}
\lambda_{4,\rm eff}(q^2;\mu)
=
\lambda_4(\mu)
+\delta\lambda_4
-\frac{\kappa_4\kappa_H}{16\pi^2}\Delta
-\frac{\kappa_4\kappa_H}{16\pi^2}B_0^{\rm fin}(q^2;m_\phi^2,m_\phi^2;\mu)\,.
\end{equation}
In the $\overline{\rm MS}$ scheme, the counterterm cancels the pole ($\delta\lambda_4 = \frac{\kappa_4\kappa_H}{16\pi^2} \Delta$) and the renormalised effective vertex is
\begin{equation}
\label{eq:lam4ren}
\lambda_{4,\rm eff}(q^2;\mu) = \lambda_4(\mu) - \frac{\kappa_4\kappa_H}{16\pi^2} B_0^{\rm fin}(q^2;m_\phi^2,m_\phi^2;\mu)\,.
\end{equation}

For equal internal masses, and defining $x\equiv\frac{q^2}{m_\phi^2}$, the finite part of $B_0$ is~\cite{Denner:2019vbn}
\begin{equation}
\label{eq:B0fin}
B_0^{\rm fin}(q^2;m_\phi^2,m_\phi^2;\mu) = -\ln\frac{m_\phi^2}{\mu^2} + f\!\left(x\right)\,,
\end{equation}
with $f$ given in \cref{eq:fdef}. 

The vertex $\lambda_{4,\rm eff}$ is a physical amplitude and cannot depend on $\mu$. Requiring this of \cref{eq:lam4ren}, with the $\mu$-dependence of $B_0^{\rm fin}$ in \cref{eq:B0fin}, fixes the beta function of $\lambda_4(\mu)$ at this order,
\begin{equation}
\label{eq:lam4beta}
\beta_{\lambda_4}
\equiv
\frac{d\lambda_4(\mu)}{d\ln\mu}
=
\frac{\kappa_4\kappa_H}{8\pi^2}\,,
\end{equation}
which integrates, with the boundary condition \cref{eq:xchannelcondition}, to
\begin{equation}
\label{eq:lam4running}
\lambda_4(\mu)
=
\frac{\kappa_4\kappa_H}{8\pi^2}
\ln\frac{\mu}{\mu_0}\,.
\end{equation}
Substituting \cref{eq:lam4running} into \cref{eq:lam4ren} and using \cref{eq:B0fin},
\begin{equation}
\label{eq:lam4eff_mu}
\lambda_{4,\rm eff}(q^2;\mu_0)
=
\frac{\kappa_4\kappa_H}{8\pi^2}\ln\frac{\mu}{\mu_0}
-\frac{\kappa_4\kappa_H}{16\pi^2}\left[2\ln\frac{\mu}{m_\phi}+f(x)\right]
=
-\frac{\kappa_4\kappa_H}{16\pi^2}\left[f(x)+2\ln\frac{\mu_0}{m_\phi}\right]\,.
\end{equation}
Using $\mu_0=c\,m_\phi$ gives
\begin{equation}
\lambda_{4,\rm eff}(q^2;c) = -\frac{\kappa_4\kappa_H}{16\pi^2} \left[ f\!\left(\frac{q^2}{m_\phi^2}\right)+2\ln c
\right],
\end{equation}
which reproduces \cref{eq:lam4eff}. For $c=1$, the form factor vanishes at $q^2=0$, as required by the $x$-channel condition.

\section{Loop amplitudes for relic density}
\label{app:relicloops}

The loop-induced $X_sX_s$ annihilation amplitudes of \cref{fig:DMannLoop} have been generated with {\sc FeynArts} and {\sc FormCalc} packages~\cite{Hahn:2000kx,Hahn:1998yk} and reduced to the scalar Passarino-Veltman functions $B_0$, $C_0$, evaluated numerically with {\sc LoopTools}~\cite{Hahn:1998yk}. Here $s=(p_1+p_2)^2$, through which $m_{X_s}$ enters the amplitudes, $m_\phi\equiv m_{\phi^{+(+)}}$ is the mass of the charged scalar in the loop and $Q_\phi$ its electric charge. 
Since $\phi$ is an $SU(2)_L$ singlet ($T_3=0$), its photon and $Z$ couplings are $g_{\gamma\phi\phi}=-Q_\phi e$ and $g_{Z\phi\phi}=-Q_\phi e\,s_w/c_w$, and the amplitudes hold for general $Q_\phi$.

The only UV-divergent bubble is the one attached to a Higgs boson. It enters through $\lambda_{4,\rm eff}(q^2;\mu_0)$ of \cref{eq:lam4eff}, with $\kappa_4\kappa_HB_0(q^2;m_\phi^2,m_\phi^2;\mu_0)\to-16\pi^2\lambda_{4,\rm eff}(q^2;\mu_0)$ for $\lambda_4(\mu_0)=0$, and all other combinations of $B_0$ functions below are UV finite.

\paragraph{Cross sections.} For a two-body final state $ab$ the cross section is
\begin{equation}
\sigma(X_sX_s\to ab)=\frac{1}{16\pi s}\,\frac{\beta_{ab}}{\beta_X}\,\overline{|\mathcal M|^2}\,S_{ab},\qquad
\beta_{ab}=\frac{\lambda_K^{1/2}(s,m_a^2,m_b^2)}{s},\quad \beta_X=\sqrt{1-\frac{4m_{X_s}^2}{s}},
\end{equation}
where $\lambda_K$ is the K\"all\'en function, $\overline{|\mathcal M|^2}$ is the squared amplitude summed over the final-state polarisations and colours and averaged over the scattering angle, $S_{ab}=1/2$ for the identical final states $\gamma\gamma$, $ZZ$, $HH$ and $S_{ab}=1$ otherwise. 

\paragraph{$\gamma\gamma$ channel.}
\begin{equation}
\label{eq:MggXsXs}
\mathcal M(X_sX_s\to\gamma\gamma) = -\frac{Q_\phi^2\,\kappa_4\alpha}{2\pi}\left[1+2m_\phi^2\,C_0(0,s,0;m_\phi^2,m_\phi^2,m_\phi^2)\right]\varepsilon_3\!\cdot\!\varepsilon_4 ,
\end{equation}
where $\varepsilon_{3,4}$ are the photon polarisation vectors.

\paragraph{$\gamma Z$ channel.}
\begin{equation}
\mathcal M(X_sX_s\to\gamma Z) = \frac{Q_\phi^2\,\alpha\,\kappa_4\, s_w}{2c_w\,\pi}\,\frac{N}{(m_Z^2-s)^2}\left[(m_Z^2-s)\,\varepsilon_3\!\cdot\!\varepsilon_4+2(\varepsilon_3\!\cdot k_4)(\varepsilon_4\!\cdot k_3)\right],
\end{equation}
with
\begin{equation}
N\equiv m_Z^2\big[B_0(m_Z^2;m_\phi^2,m_\phi^2)-B_0(s;m_\phi^2,m_\phi^2)\big]+(m_Z^2-s)\big[1+2m_\phi^2\,C_0(0,s,m_Z^2;m_\phi^2,m_\phi^2,m_\phi^2)\big],
\end{equation}
and $k_{3,4}$, $\varepsilon_{3,4}$ the momenta and polarisation vectors of the $\gamma$ and $Z$ respectively.

\paragraph{$ZZ$ channel.} With $B_0(m_Z^2)\equiv B_0(m_Z^2;m_\phi^2,m_\phi^2)$, $B_0(s)\equiv B_0(s;m_\phi^2,m_\phi^2)$, $C_0\equiv C_0(m_Z^2,s,m_Z^2;m_\phi^2,m_\phi^2,m_\phi^2)$, and $T_H$ carrying the $\lambda_{4,\rm eff}$ substitution above,
\begin{align}
T_1 &= \frac{8\,Q_\phi^2\,s_w^2}{4m_Z^2-s}\Big[m_Z^2\big(B_0(m_Z^2)-B_0(s)\big)-\big(m_Z^4-m_\phi^2(4m_Z^2-s)\big)C_0\Big],\notag\\[4pt]
T_H &= -\frac{16\pi^2 v^2}{\kappa_4\,s_w^2}\,\frac{\lambda_{4,\rm eff}(s)}{s-m_h^2+im_h\Gamma_h} ,\notag\\[4pt]
G &= \alpha\Big[4\,Q_\phi^2\,s_w^2+T_1+T_H\Big],\notag\\[4pt]
K &= -\frac{8\,Q_\phi^2\,\alpha\, s_w^2}{s(s-4m_Z^2)^2}\Big[(8m_Z^4-6m_Z^2 s+s^2)\notag\\
&\qquad-2m_Z^2(2m_Z^2+s)\big(B_0(m_Z^2)-B_0(s)\big)\notag\\
&\qquad+2\big(2m_Z^4(s-m_Z^2)+m_\phi^2(8m_Z^4-6m_Z^2s+s^2)\big)C_0\Big],
\end{align}
the amplitude is
\begin{equation}
\mathcal M(X_sX_s\to ZZ) = -\frac{\kappa_4}{8c_w^2\pi}\Big[G\,\varepsilon_3\!\cdot\!\varepsilon_4 + K\,(\varepsilon_3\!\cdot k_4)(\varepsilon_4\!\cdot k_3)\Big] ,
\end{equation}
with $\varepsilon_{3,4},k_{3,4}$ the polarisation vectors and momenta of the two final-state $Z$ bosons.

\paragraph{$Zh$ channel.} The amplitude is proportional to $B_0(s)+2B_1(s)$, which vanishes for equal internal masses, so the $Zh$ final state does not contribute.

\paragraph{$WW$ channel.} Purely Higgs-mediated, with the $\lambda_{4,\rm eff}$ substitution above,
\begin{equation}
\mathcal M(X_sX_s\to WW)=\frac{2\pi\alpha\, v^2}{s_w^2}\,\lambda_{4,\rm eff}(s)\,\frac{\varepsilon_3\!\cdot\!\varepsilon_4}{s-m_h^2+im_h\Gamma_h} ,
\end{equation}
proportional to $\varepsilon_3\!\cdot\!\varepsilon_4$ only, as the amplitude is a scalar exchange.

\paragraph{$HH$ channel.} With $C_0^{HH}\equiv C_0(m_h^2,s,m_h^2;m_\phi^2,m_\phi^2,m_\phi^2)$ for on-shell Higgs bosons, $k_1^2=k_2^2=m_h^2$,
\begin{equation}
\label{eq:MXsXsHH}
\mathcal M(X_sX_s\to HH)=-\lambda_{4,\rm eff}(s)\left[1+\frac{3m_h^2}{s-m_h^2+im_h\Gamma_h}\right]-\frac{\kappa_4\kappa_H^2 v^2}{8\pi^2}C_0^{HH}\;.
\end{equation}
The first term collects the contact bubble and the $s$-channel Higgs exchange through the SM trilinear coupling. The second is the triangle with two single-Higgs vertices, summed over the two orientations of the loop.

\paragraph{$f\bar f$ channels.} For a Dirac fermion $f$ of mass $m_f$, quark or lepton, the Higgs-mediated amplitude is
\begin{equation}
\mathcal M(X_sX_s\to f\bar f) = \frac{m_f}{s-m_h^2+im_h\Gamma_h}\,\lambda_{4,\rm eff}(s)\;\bar u(k_3)v(k_4).
\end{equation}
The colour factor $N_c=3$ multiplies the quark channels.

\paragraph{Lepton pairs.} The $\phi$ loop contributes to $X_sX_s\to\ell_a^+\ell_b^-$ through a triangle with two $\phi$ propagators and an internal fermion $k$ of mass $m_k$ between the two lepton vertices, with couplings $c_{ak}$. For $\phi^+$, $c_{ak}=\lambda_{ak}$ and $k$ runs over the neutrinos, with $m_k=0$. For $\phi^{++}$, $c_{ak}=2f_{ak}$ and $k$ runs over the charged leptons: the vertex $\phi^{++}\ell_a^+\ell_k^+$ is $2f_{ak}P_R$ for every $(a,k)$, from the two contractions of $\bar\ell^c_a\ell_k$ with the symmetric $f_{ak}$, and charge conservation fixes the final state to be opposite-sign. For $a=b=l$,
\begin{align}
\mathcal M(X_sX_s\to\ell_l^+\ell_l^-)&=\frac{m_l\,\lambda_{4,\rm eff}(s)}{s-m_h^2+im_h\Gamma_h}\,\bar u(k_3)v(k_4)\notag\\
&\quad+\frac{\kappa_4\,m_l}{16\pi^2(s-4m_l^2)}\sum_k c_{lk}^2\Big[B_0(m_l^2;m_k^2,m_\phi^2)-B_0(s;m_\phi^2,m_\phi^2)\notag\\
&\qquad\qquad-(m_l^2+m_k^2-m_\phi^2)\,C_0(m_l^2,s,m_l^2;m_k^2,m_\phi^2,m_\phi^2)\Big]\bar u(k_3)v(k_4)\,.
\end{align}
The first term is the Higgs-mediated amplitude of the $f\bar f$ channel. The loop amplitude is proportional to $m_l$ because the chiral couplings of $\phi$ forbid a mass insertion on the internal line, and the helicity flip occurs on the external legs. For $a\neq b$ there is no Higgs-mediated term, and the amplitude has the same triangle structure, $\sum_kc_{ak}c_{bk}\,\mathcal A_k(s;m_a,m_b,m_k,m_\phi)$, built from $B_0$, $C_0$ and $\lambda_K(s,m_a^2,m_b^2)$. Its explicit form is lengthy and is omitted.

%%%%%%%%%%%%%%%%%%%%%%%%%%%%%%%%%%%%%%%%%%%%%%%%%%%%%%%%%%%%%%%%%%%%%%%%%%%%%%
\bibliographystyle{JHEP}
\bibliography{biblio}

\end{document}